\documentclass[graybox,natbib,nosecnum]{svmult}
\bibpunct{(}{)}{;}{a}{}{,} % suppress commas between author-names and year

\pdfoutput=1   %forces use of pdflatex. Disable if you prefer to use .eps or .ps figures.
\usepackage{mathptmx}       % selects Times Roman as basic font
\usepackage{helvet}         % selects Helvetica as sans-serif font
\usepackage{courier}        % selects Courier as typewriter font
\usepackage{type1cm}        % activate if the above 3 fonts are
\usepackage{makeidx}         % allows index generation
\usepackage{graphicx}        % standard LaTeX graphics tool
\usepackage{multicol}        % used for the two-column index
\usepackage[bottom]{footmisc}% places footnotes at page bottom
\usepackage[normalem]{ulem}	% for strike-through of text with \sout{}  
\usepackage{hyperref}  %for hyperlinks

\usepackage{amssymb}
\usepackage{amsmath}
\usepackage{xcolor}

\usepackage{soul}   % for high-lighting of text
\usepackage{amssymb} 
\usepackage{textcomp} % \textdiv

\newcommand\restr[2]{{% we make the whole thing an ordinary symbol
  \left.\kern-\nulldelimiterspace % automatically resize the bar with \right
  #1 % the function
  \vphantom{\big|} % pretend it's a little taller at normal size
  \right|_{#2} % this is the delimiter
  }}

\makeindex             % used for the subject index
\begin{document}

\title*{Exoplanet Occurrence Rates from Microlensing Surveys}
\author{Przemek Mr\'oz and Rados\l{}aw Poleski}
\institute{Przemek Mr\'oz \at Astronomical Observatory, University of Warsaw, Al.~Ujazdowskie 4, 00-478 Warszawa, Poland, \email{pmroz@astrouw.edu.pl}
\and Rados\l{}aw Poleski \at Astronomical Observatory, University of Warsaw, Al.~Ujazdowskie 4, 00-478 Warszawa, Poland, \email{rpoleski@astrouw.edu.pl}
\and Authors contributed equally.}

\maketitle

\abstract{
The number of exoplanets detected using gravitational microlensing technique is currently larger than 200, which enables population studies. Microlensing is uniquely sensitive to low-mass planets orbiting at separations of several astronomical units, a parameter space that is not accessible to other planet-detection techniques, as well as free-floating planets, not orbiting around any star. In this review, we present the state-of-the-art knowledge on the demographics of exoplanets detected with microlensing, with a particular emphasis on their occurrence rates. We also summarize the current knowledge about free-floating planets, an elusive population of objects that seem to be more common than ordinary, gravitationally bound exoplanets.
}

\section{Introduction}

\begin{figure}[t]
\includegraphics[scale=.65]{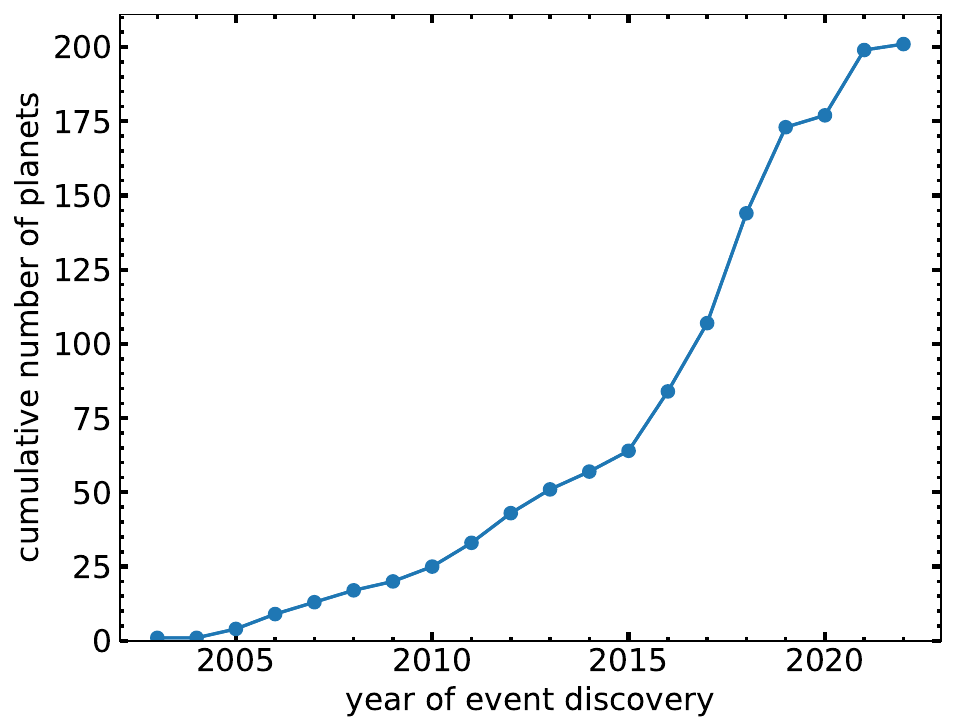}
\caption{Cumulative number of known extrasolar planets detected with gravitational microlensing as a function of the event discovery year. The number of planets detected per year increases in 2011 and 2016, when large-scale surveys started: OGLE-IV and KMTNet, respectively. The smaller number of planets in the last few years is caused by the telescopes shutdown due to pandemic (2020) and the publication bias (2021 and 2022). Data were taken from the NASA Exoplanet Archive (accessed 09/08/2023).}
\label{fig:number}
\end{figure}

Since the inception of an idea that extrasolar planets may be detected with gravitational microlensing more than 30 years ago by \citet{mao1991} and \citet{gould1992}, microlensing surveys have discovered over 200 exoplanets (as shown in Fig.~\ref{fig:number}), and this number is expected to substantially increase in the upcoming years. Microlensing is unique among exoplanet-detection techniques, as it enables us to discover cold low-mass planets orbiting far from their host stars, which are largely inaccessible to other planet-detection methods. Microlensing is also unique because it is a transient phenomenon; we can infer properties of the planet from the available data, but the opportunities for the follow-up observations are restricted. Therefore, microlensing surveys for exoplanets are especially suited for studying their demographics, most importantly, measuring the exoplanet occurrence rates.

Gravitational microlensing is the effect predicted by Einstein's theory of general relativity. It occurs when light from a distant background star (called ``source'') is deflected by the gravitational field of an intervening object (called ``lens''). If the lens is an isolated body, usually two images of the source star are created. These images cannot normally be resolved with traditional telescopes, but their combined brightness is larger than the brightness of the source when unlensed (i.e., the source is magnified). The relative motion of the observer, the lens, and the source causes the total magnification to vary in time, producing a characteristic light curve of a microlensing event \citep{paczynski1986}.

\begin{figure}[t]
\centering
\includegraphics[width=0.9\textwidth]{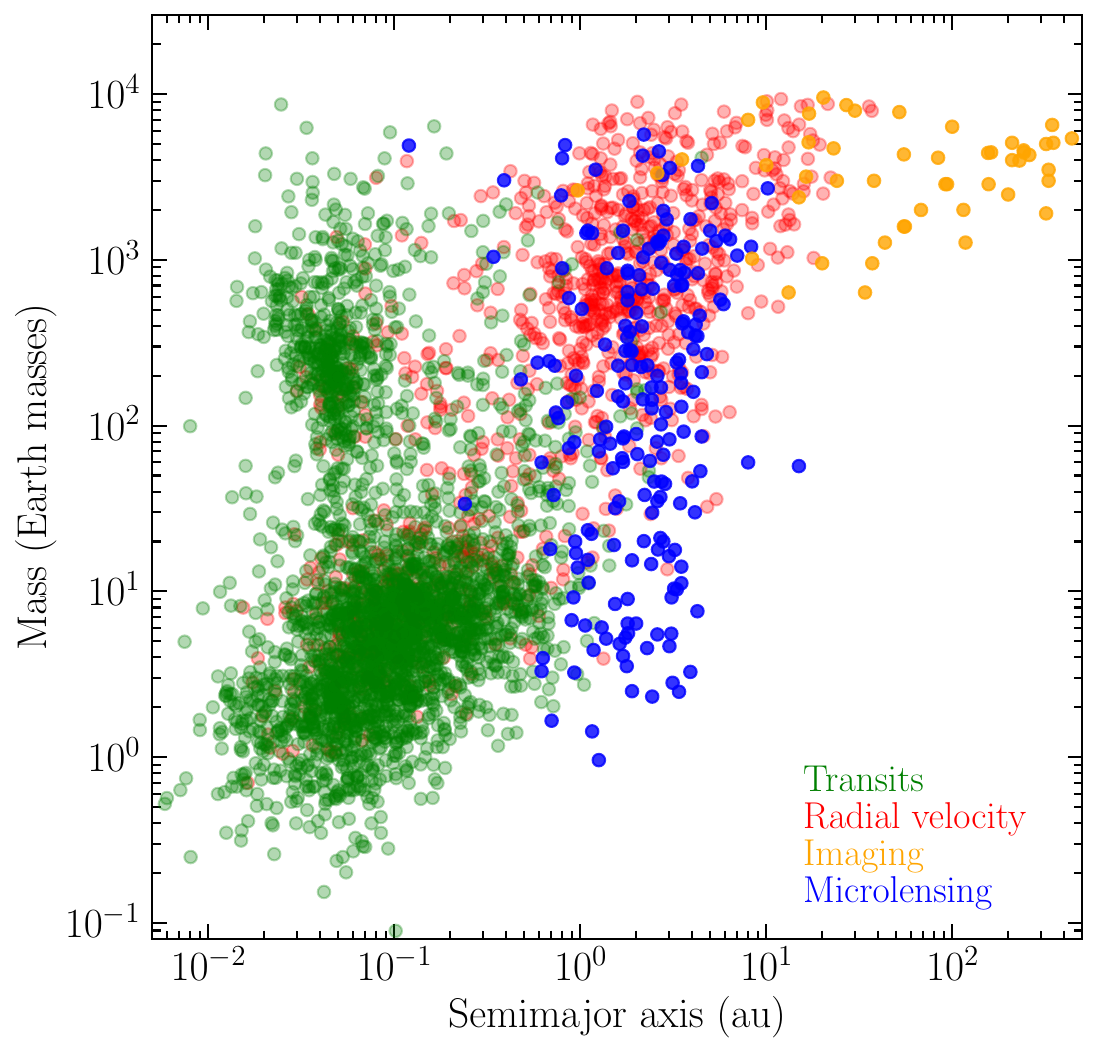}
\caption{Properties of known exoplanets in mass vs.~semi-major axis plane. Color codes the detection technique: green---transits, red---radial velocity, orange---direct imaging, blue---microlensing. Data were taken from the NASA Exoplanet Archive (accessed 09/08/2023).}
\label{fig:penny_plot}
\end{figure}

If the lens is orbited by a planet and the planet happens to be located in the vicinity of one of the images created by microlensing, the planet may induce a short-lasting anomaly in an otherwise smooth light curve of the event. 
By modeling the shape of the light curve, it is possible to measure two parameters that describe the planet-star system: $q$ -- the planet to star mass ratio and $s$ -- the projected separation relative to the angular Einstein ring radius $\theta_\mathrm{E}$. The latter quantity defines the characteristic angular scale of the event and depends on the mass of the lens $M$ and the relative lens-source parallax $\pi_{\rm rel}$:
\begin{equation}
    \theta_{\rm E} = \sqrt{\kappa M \pi_{\rm rel}} = 0.5\,\mathrm{mas}\left(\frac{M}{0.3\,M_{\odot}}\right)^{1/2}\left(\frac{\pi_{\rm rel}}{0.1\,\mathrm{mas}}\right)^{1/2},
\end{equation}
where $\kappa = 4G/c^2\mathrm{au} = 8.144$\,mas\,$M_{\odot}^{-1}$ is a constant and $\pi_{\rm rel}=\mathrm{au}/D_{\rm l}-\mathrm{au}/D_{\rm s}$, where $D_{\rm l}$ and $D_{\rm s}$ are distances to the lens and the source, respectively.
Microlensing is most sensitive to planets located near the Einstein ring, that is, at a separation of between 1 and 5\,au (Fig.~\ref{fig:penny_plot}) for typical configurations in the Milky Way. The peak sensitivity coincides with the location of the snow line, a region in the protoplanetary disk where the majority of planets are thought to be formed \citep[e.g.,][]{ida2004}.

Additional information is required for translating parameters of the microlensing model $(s,q)$ to physical parameters of the system, such as the projected star--planet separation $a_{\perp}$ or the mass of the planet $m_{\rm p}$. 
This information can be obtained from the so-called second-order effects in microlensing light curves. The finite-source effects \citep{gould1994,nemiroff1994,witt1994}, which are detectable in the most of planetary microlensing events, enable one to measure the angular Einstein radius (see \citet{yoo2004} for the practical implementation of the method). The microlensing parallax effects \citep{gould1992b} yield the parameter $\pi_{\rm E}\equiv \pi_{\rm rel}/\theta_{\rm E}$. The microlensing parallax can be measured in either long-duration events, in which the orbital motion of the Earth around the Sun induces small deviations in the microlensing light curve, or events observed simultaneously from two separated observatories. Then, the mass of the lens can be estimated directly:
\begin{equation}
    M = \frac{\theta_{\rm E}}{\kappa \pi_{\rm E}},
\end{equation}
and the mass of the planet is simply
\begin{equation}
    m_{\rm p} = \frac{qM}{1+q}.
\end{equation}
The distance to the planetary system can be inferred based on definitions of $\pi_\mathrm{rel}$ and $\pi_\mathrm{E}$:
\begin{equation}
    \frac{\mathrm{au}}{D_{\rm l}} = \theta_{\rm E}\pi_{\rm E}+\frac{\mathrm{au}}{D_{\rm s}}.
\end{equation}
Additionally, the mass and distance to the lens can sometimes be inferred when the light from the lens (host star) is directly detected in high-resolution images taken several years after the event, when the lens and the source separate in the sky \citep[e.g.,][]{alcock2001,bennett2015,batista2015}. 

Once we know the distance to the lens, the value of $s$ can be translated to a 2D separation in physical units $a_\perp = s D_{\rm l}\theta_\mathrm{E}$. The deprojection of $a_\perp$ from a 2D separation to a full 3D separation is even more complicated. It can be done in rare cases of triple lenses (one planet plus two stars or two planets orbiting one star) using orbital stability considerations \citep{madsen2019}. An alternative deprojection method is to measure the orbital motion of the planet during the event with an accuracy high enough that the full Keplerian orbit has to considered. So far, this technique has been used only in the case of star-star lenses \citep[e.g.,][]{skowron2011,wyrzykowski2020}, which light curves show longer anomalies. A de-projection method that can be always applied is to statistically guess the deprojection factor based on the assumption of a random position of the planet on its orbit, but this approach has a limited value.

We refer the reader to reviews by \citet{gaudi2012}, \citet{gould2016rev}, and \citet{batista2018} for a detailed description of the phenomenology of planetary microlensing events and the methods used to characterize the exoplanets detected.

\section{Microlensing surveys}

Modern experiments looking for gravitational microlensing events target mostly the dense regions of the Galactic bulge, where the chances of observing microlensing are highest. These projects repeatedly image large areas around the Galactic center, measuring the brightness of hundreds of millions stars at a cadence as short as several minutes. The Optical Gravitational Lensing Experiment (OGLE) is the longest-running microlensing survey, it has been observing the sky since 1992. The project is currently in its fourth phase (since 2010), in which it uses a mosaic camera with a field of view of 1.4\,deg$^2$, mounted on a 1.3-m Warsaw Telescope, located at the Las Campanas Observatory, Chile \citep{udalski2015b}. The Microlensing Observations in Astrophysics (MOA) \citep{sumi2003} carries out a microlensing survey with a 1.8-m MOA-II telescope, which has a 2.18\,deg$^2$ field of view, located at the Mt.~John University Observatory, New Zealand. The newest survey, the Korea Microlensing Telescope Network (KMTNet; \citealt{kim2016}), uses three identical 1.8-m telescopes, each equipped with a $2^{\circ}\times 2^{\circ}$ camera. The KMTNet telescopes are located at the Cerro Tololo Inter-American Observatory (Chile), the Siding Spring Observatory (Australia), and the South African Astronomical Observatory (South Africa).
Apart from these dedicated surveys, an increasing number of microlensing events (including planetary events) is detected with large all-sky surveys, such as the \textit{Gaia} satellite \citep{wyrzykowski2023,wu2023}, the Zwicky Transient Facility \citep{rodriguez2022,medford2023}, or the All Sky Automated Survey for SuperNovae (ASAS-SN; e.g., \citealt{fukui2019}).

\section{Occurrence Rates of Bound Planets}

\subsection{Theory of microlensing occurrence rates}

The intrinsic properties of lensing planetary systems are typically derived based on the $(s, q)$ parameters of detected planets and the detection efficiency of the survey from which these detections come from. The detection efficiency of a survey $S$ as a function of $(s, q)$ is defined as the expected number of planets to be discovered if every lensing star had a planet with a given $(s, q)$. 

There are two general approaches for estimating detection efficiency for a single event, which were presented by \citet{gaudi00} and \citet{rhie00}. The latter method is much more frequently used than the former one. 
Similar to the analysis of planetary events, one has to assume what is the angle between the projected binary axis and the trajectory of the source ($\alpha$).
For a given event and $(s, q, \alpha)$, a synthetic light curve is created with the same times of observations and the same noise properties as the original light curve. 
In the method of \citet{gaudi00}, a binary-lens model (with fixed values of $s$, $q$, and $\alpha$) is fitted to the synthetic data, and the difference $\Delta\chi^2$ is calculated between the best-fit binary-lens model and the best-fit single-lens model. In the method of \citet{rhie00}, a single-lens model is fitted to the synthetic data, and the difference $\Delta\chi^2$ is calculated between the best-fit single-lens model to the synthetic data and the best-fit single-lens model to the original data. The anomaly is considered as detected if $\Delta\chi^2$ is greater than some threshold value and other conditions are met (e.g., the number of anomalous epochs is larger than a threshold value). 
The $\Delta\chi^2$ threshold used by different authors varies between 60 and 500 \citep[see references in][]{udalski18}.

The above strategy allows checking if the planet would be detected for given $(s, q, \alpha)$ parameters. To obtain the detection efficiency for the event $i$ and a given pair $(s, q)$, i.e., $S_i(s, q)$, one simulates a grid of uniformly spaced $\alpha$ values and sets $S_i(s,q)$ to be the number of possible detections divided by the number of the grid points. Then one obtains the survey detection efficiency by summing the detection efficiencies over all the events analyzed: $S(s, q) = \sum_i S_i(s,q)$.

An important aspect of the detection efficiency calculations is the assumed value of the normalized source radius: 
\begin{equation}\label{eq:rho1}
\rho = \frac{\theta_*}{\theta_{\rm E}} = 0.013 \left(\frac{\theta_{*}}{6.3\,\mu\rm{as}}\right) \left(\frac{M}{0.3\,M_{\odot}}\right)^{-1/2} \left(\frac{\pi_{\rm rel}}{0.1\,\rm{mas}}\right)^{-1/2},
\end{equation}
where $6.3\,\mu\mathrm{as}$ is the angular radius of a red clump star (i.e., $11\,R_{\odot}$) located in the Galactic center. The value of $\rho$ can significantly affect almost all light curves in which planets can be detected, yet, this parameter is rarely constrained for single-lens events, which are used for detection efficiency calculations. If the finite-source effects are detected, then a smaller $\rho$ generally results in a higher amplitude and a shorter duration of the anomaly than a larger $\rho$.

So far, the approaches used to estimate $\rho$ varied significantly in assumptions used. The source brightness derived from the light curve fit can be combined with the angular Einstein radius $\theta_\mathrm{E}$ expected from a Galactic density and kinematics model (based on the event position and timescale $t_\mathrm{E}$). Some authors have used the most likely output from this procedure and calculated $S_i(s,q)$ assuming that value of $\rho$ \citep{gaudi2002}. A more advanced approach is to take into account the uncertainty in $\rho$, which was done either by using three values: $\rho-\sigma_\rho$, $\rho$, and $\rho+\sigma_\rho$ \citep[e.g.,][]{suzuki2016}, or taking at least a few samples and weighting them using the importance sampling technique \citep[ten samples in the case of][]{poleski21}. Both above approaches sample $\rho$ by independently sampling values of $\theta_*$ and $\theta_\mathrm{E}$ and then calculating their ratio. In rare cases, such procedures lead to the results that are not self-consistent: it may happen that the calculated $\rho$ is large enough for finite source effects to be visible, even though this effect is not seen in the light curve. Other problem may be the lens flux that is brighter than the observed unmagnified flux. In order to overcome these issues, one could fit the light curve together with baseline fluxes using the model that has all important physical parameters ($D_{\rm l}$, $D_{\rm s}$, $M$ etc.) in addition to the parameters of the microlensing light curve. Such a fit can be done using Bayesian approach and priors on many parameters from Galactic models, stellar isochrones, and color--surface brightness relations. 

The final result of the considerations presented here is the occurrence rate defined as a number of planets per star per dex of $s$ and $q$, i.e.,
\begin{equation}\label{eq:f}
f(s, q) = \frac{d^2\,N}{d\,\log{s}\,d\,\log{q}}.
\end{equation}
Similarly, the occurrence rate can be defined for other variables, for example, $f(q) = d\,N/d\,\log{q}$.

The occurrence rate $f(s,q)$ is derived from the properties of the sample of $N_\mathrm{obs}$ detected planets $\{s_i,q_i\}$ and the survey detection efficiency. We parameterize $f(s,q)$ using a vector of parameters $\mathbf{p}$ (for example, power-law exponents) and derive these parameters using the Bayesian techniques and likelihood (i.e., the probability of obtaining a set of $\{s_i,q_i\}$ data given parameters $\mathbf{p}$) defined as:
\begin{equation}\label{eq:l}
\mathcal{L}(\{s_i,q_i\};\mathbf{p}) = e^{-N_\mathrm{exp}(\mathbf{p})} \prod^{N_\mathrm{obs}}_{i=1}f(s_i, q_i; \mathbf{p})S(s_i, q_i),
\end{equation}
where $N_\mathrm{exp}(\mathbf{p})$ is the number of expected detections:
\begin{equation}
N_\mathrm{exp}(\mathbf{p}) = \int f(s, q; \mathbf{p})S(s, q) d\,\log{s}\,d\,\log{q}
\end{equation}
and the integration is done over the same domain of $(s, q)$ as searched for planets. The Eq.~(\ref{eq:l}) can be derived by dividing the domain of $f(s, q)$ into bins, assuming the number of planets detected in each bin follows the Poisson distribution, and calculating the limiting case of infinitesimally small bins (i.e., bins with either zero or one planet detected). 

The studies aimed at deriving the planet frequency first presented the detection efficiency for single events \citep[e.g.,][]{rhie00,kubas2008}. Later, the detection efficiency was studied for a statistically significant number of events \citep{gaudi2002,snodgrass2004,tsapras16}. Finally, the microlensing planet occurrence rate was first estimated quantitatively by \citet{sumi10}. They used ten microlensing planets known at that time and detection efficiency estimated (but not derived in detail) to be $S(q) \propto q^{0.6\pm0.1}$. Based on these pieces of information, only the power law index of the mass-ratio function was derived: $dN/d\log{q} \propto q^{-0.68\pm0.20}$. Because of the simplified assumptions about the detection efficiency, the normalization of the mass-ratio function could not be derived.

We present planet occurrence rates derived from microlensing data in Table~\ref{tab:bound}. Some of the studies presented there are also described in more detail below. Note that different studies present occurrence rates as a function of different variables and the usage of results from previous investigations.

\begin{table}
\caption{Bound planet occurrence rates}
\label{tab:bound}
\begin{tabular}{l l l l l}
\hline\noalign{\smallskip}
Reference & $n_\mathrm{pl}$ $^a$ & $n_\mathrm{ev}$ $^b$ & Result$^c$ & Comments \\
\noalign{\smallskip}\svhline\noalign{\smallskip}
\citet{sumi10}        & 10 & -- & $f(q) \propto q^{-0.68\pm0.20}$ & \\
\citet{gould10}       & 6 & 13 & $f(s, q) = 0.36\pm0.15$ & for $q=5\times10^{-3}$ \\
\citet{cassan12}      & 3 & 440 & $f(a, m_{\rm p}) = 10^{-0.62\pm0.22}\times$ &  \\
 & & & $\left(\frac{m_{\rm p}}{M_\mathrm{Sat}}\right)^{-0.73\pm0.17}$ & \\
\citet{shvartzvald16} & 9 & 224 & $f(q) \propto q^{-0.50\pm0.17}$ & for $q<10^{-1.4}$; the slope is \\
 & & & & $0.32\pm0.38$ for larger $q$  \\
\citet{suzuki2016}      & 22 & 1474 & $f(s,q) = 0.61^{+0.21}_{-0.16} \times$ & $q$ exponent changes \\
 & & & $s^{0.49\pm0.48}\left(\frac{q}{1.7\times10^{-4}}\right)^{-0.93\pm0.13}$ &  to $0.6^{+0.5}_{-0.4}$ for $q<q_\mathrm{br}$ \\
\citet{udalski18} & 7 & -- & $f(q) \propto q^{1.05^{+0.78}_{-0.68}}$ & for $q < 10^{-4}$ \\
\citet{poleski21}     & 5 & 3095  & $f(s,q)= 1.04^{+0.78}_{-0.57} \times $ & only $2<s<6$ \\
 & & & $s^{1.09\pm0.64}\left(\frac{q}{1.7\times10^{-4}}\right)^{-1.15\pm0.25}$ & \\
\citet{koshimoto2021}   & 28 & -- & $f(R_L,M) \propto R_L^{0.2\pm0.4}M^{0.7^{+0.8}_{-0.6}}$ & $R_L$ is the Galactocentric \\
 & & & & lens distance \\
\noalign{\smallskip}\hline\noalign{\smallskip}
\end{tabular}

$^a$ -- number of planets detected; 
$^b$ -- number of events analyzed; 
$^c$ -- for definition of the planet occurrence rate, see Eq.~\ref{eq:f}.
\end{table}

\subsection{Occurrence rates from high-magnification events}

The planet occurrence rate can be derived either from a large number of events with a whole range of magnifications or from a much smaller number of high-magnification events. The peaks of high-magnification events are the most important because they give high chances of detecting planets, if the latter were present \citep{griest98}. The high-magnification events have two additional advantages: they are bright at the peak and can be selected a day or so before the peak. These features reduce the amount of telescope time needed to discover a planet. High-magnification events often get bright enough to be observed by a small-aperture telescopes and even amateur astronomers may collect scientifically useful photometry. Such an experiment was conducted by the Microlensing Follow Up Network ($\mu$FUN) and its results were presented by \citet{gould10}. They analyzed data from a four-year-long experiment which densely followed-up 13 events with a peak magnification greater than 200. Six planets in five planetary events were detected in this experiment, which allowed the authors to measure the planet mass-ratio function:
\begin{equation}
\restr{f(s, q)}{q=5\times10^{-3}} = 0.36\pm0.15\,\mathrm{dex}^{-2}.
\end{equation}

A larger sample of events was studied in a similar manner by \citet{cassan12}. They used photometry collected by the Probing Lensing Anomalies Network (PLANET) collaboration. The planet detection efficiency was calculated for the $(s, q)$ parameters and scaled to $(a_\perp, m_p)$. This scaling was done separately for each event based on the Einstein timescale ($t_\mathrm{E}$) measured. Then, the projected separation $a_\perp$ was further scaled to the semi-major axis $a$ assuming isotropic circular orbits \citep{kubas2008}. The PLANET collaboration detected three planets and used earlier studies \citep{sumi10,gould10} as additional constraints. The result of the \citet{cassan12} investigation was:
\begin{equation}
f(a, m_{\rm p}) = 10^{-0.62\pm0.22}\left(\frac{m_{\rm p}}{M_\mathrm{Sat}}\right)^{-0.73\pm0.17},
\end{equation}
where $M_\mathrm{Sat} = 95\,M_{\oplus}$ is the mass of Saturn. This planet occurrence rate was integrated over the semi-major axis range $0.5 \div 10\,\mathrm{au}$ for Jupiter-mass ($0.3\div 10\,M_{\rm J}$), cool Neptunes ($10\div 30~M_\oplus$), and super-Earth planets ($5\div 10\,M_\oplus$) resulting in 
$17^{+6}_{-9}\%$, $52^{+22}_{-29}\%$, and $62^{+32}_{-37}\%$, respectively. These numbers indicated that planets frequently orbit stars in the Milky Way.

The results obtained by \citet{cassan12} were assumed as a baseline for the predictions of the planet yield by \citet{penny2019} in their preparation of the microlensing survey for the \textit{Nancy Grace Roman Space Telescope} (formerly \textit{WFIRST}). The \textit{Roman} telescope is currently predicted to be launched in 2026 or 2027, hence, it is likely that the \citet{cassan12} result will remain a reference for detailed preparations of the \textit{Roman} microlensing observations and studies of results of these observations.

The above studies of microlensing planet occurrence rate were followed-up by studies that aimed at more detailed understanding of the whole population of planets. A joint analysis of microlensing and radial velocity results was conducted by \citet{clanton2014a,clanton2014b}. They concluded that the statistical results from both methods agree in the part of parameter space where their sensitivity overlaps. Furthermore, a single planetary occurrence rate was found consistent with constraints from microlensing, radial velocity, and direct imaging techniques \citep{clanton2016}.

\subsection{Occurrence rates from modern microlensing surveys}

The most detailed study of bound microlensing planet statistics was done based on the MOA survey data and presented by \citet{suzuki2016}. They used data collected between 2007 and 2012: 1474 events with 23 planets detected. These are planets detected from the MOA data alone. For each of these planets, the parameters used in the statistical analysis were taken from the fit to the light curve that combines all survey and follow-up data available. To analyze the occurrence rate, \citet{suzuki2016} first plotted the histogram of $q_i/S(q_i)$, i.e., the number of detected planets corrected for survey sensitivity, where $S(q)=\int S(s,q)d\,s$. This histogram has a constant bin width in $\log{q}$ and an approximately constant slope for $\log{q} > -4.5$. There is no planet with a lower $q$. The shape of the histogram combined with a power-law shape of $S(q)$ points to a planet occurrence rate that is a broken power law of $q$, i.e.,
\begin{equation}
f(s,q; A, q_\mathrm{br}, n, p, m) =
  \begin{cases}
    A\left(\frac{q}{q_\mathrm{br}}\right)^n s^m & \text{for}\quad q\ge q_\mathrm{br} \\
    A\left(\frac{q}{q_\mathrm{br}}\right)^p s^m & \text{for}\quad q<q_\mathrm{br}. \\
  \end{cases}
\end{equation}
This broken power-law is favored over the unbroken power-law by a Bayes factor of at least 21. The parameters of $f(s,q)$ were derived by jointly analyzing the MOA data with \citet{gould10} and \citet{cassan12} data. The final results are provided for an assumed fixed mass-ratio break at $q_\mathrm{br}= 1.7\times10^{-4}$. We do not see a clear argument for fixing $q_\mathrm{br}$ at this value, and if the parameter is left free in the fits, then the uncertainty of $p$ (i.e., the power-law exponent of $q$ below the break) is larger than presented in Table~\ref{tab:bound}.

\citet{suzuki2016} calculated the planet occurrence rate as a function of microlensing parameters $s$ and $q$. Translating it to the physical parameters (separation and mass) requires high-resolution imaging observations. Such observations have been collected for many events from the \citet{suzuki2016} sample, for example as a part of the NASA Keck Key Strategic Mission Support program \citep[e.g.,][]{bhattacharya2018}.

The results obtained by \citet{suzuki2016} were compared with the predictions of the core-accretion theory of planet formation by \citet{suzuki2018b}. They simulated a population of planets around lenses in all 1474 events studied by \citet{suzuki2016}; planets were simulated using approaches of two groups preparing such models: \citet{ida2004} and the Bern group \citep{mordasini2009}. The agreement between the simulated and observed populations could not be found, even though a few different assumptions were used. The discrepancies between simulated and observed populations seem most problematic for $q\approx 2\times10^{-4}$. For a typical lens mass, this mass ratio corresponds to the mass range of failed gas giant cores, i.e., $\approx20\div80\,M_\oplus$. The core-accretion model predicts that there should be very few failed gas giant cores, which is contradicted by the microlensing findings. Additionally, \citet{suzuki2018b} point that the event OGLE-2012-BLG-0950 has the lens mass of $0.58\pm0.04\,M_\odot$ and the planet mass of $39\pm8\,M_\oplus$ \citep{bhattacharya2018}, which is the kind of system that should be very rare.

The \citet{suzuki2016} analysis was followed-up by more detailed studies. \citet{udalski18} studied the mass-ratio function for planets with $q<10^{-4}$ based on all known such planets at that time, i.e., without limiting to planets found by a single survey. These planets were found using different observing approaches, which prevents using statistical methods typically used to derive the planet occurrence rate. Instead, \citet{udalski18} used a method called $V'/V'_\mathrm{max}$, which was originally developed for studies of a luminosity function of quasars \citep{schmidt1968}. In this method, one assumes the ratio of a volume density of quasars at redshift $z'$ and now $\rho(z')$ and calculates a generalized volume to each quasar detected at redshift $z$: $V'(z)=\int^z_0\rho(z')d\,V(z')$. For each quasar, one also calculates the generalized volume to the maximum redshift at which a given quasar could be discovered ($V'_\mathrm{max}$). The functional form of $\rho(z)$ can be found by requesting that $V'/V'_\mathrm{max}$ values are consistent with being drawn from a uniform distribution from 0 to 1 (i.e., the mean should be $1/2$, etc.). This method was used by \citet{udalski18} with $\rho(z)$ replaced by $f(s, q)$ and $V(z)$ replaced by $S(q)$ and resulted in $f(q) \propto q^{1.05^{+0.78}_{-0.68}}$ based on seven planets with $q < 10^{-4}$. 

The sample of 15 planets with $q<3\times 10^{-4}$ out of 16 such planets known at a time was analyzed by \citet{jung19}. They noticed that the observed distribution of mass ratios is uniform in $\log q$. Since the sensitivity to planets decreases for smaller mass ratios, this observation indicated that the intrinsic frequency of planets above the putative break must increase for smaller $q$. 
\citet{jung19} additionally made a crucial assumption that the sensitivity to planets $S(q)$ as a function of $q$ can be approximated as a simple power-law function, even below the putative break. Under that assumption, the observed distribution of $\log q$ depends on the location $q_{\rm br}$ and the ``strength'' of the mass-ratio break (that is, the change of the power-law slope). \citet{jung19} found that the sample of events they analyzed was consistent with the break at $q_{\rm br} \approx 0.55 \times 10^{-4}$, much lower than assumed by \citet{suzuki2016}. Moreover, the change of the power-law slope at the break seemed to be considerable, $\Delta n > 3.3$, indicating a substantial drop in the number of low-mass planets.

Since the location of the break inferred by \citet{jung19} was close to the lowest-$q$ planet in their sample, this immediately raised the possibility that the break may be an artifact of a sudden drop in sensitivity, not the paucity of low-mass planets, contrary to the assumption made by \citet{jung19}. To check this hypothesis, a larger sample of uniformly selected planets was needed. This sample is currently being built with the KMTNet's \textsc{AnomalyFinder} algorithm \citep{zang2021}, which is responsible for the rapid growth in the number of discovered planets seen in Fig.~\ref{fig:number}. The application of the \textsc{AnomalyFinder} algorithm to the KMTNet data collected in 2016--2019 led to the discovery of about 100 exoplanets \citep{hwang2022a, wang2022a, zang2022a, zang2023a, gould2022a, jung2022a, jung2023a, shin2023a, ryu2023a}, a statistical sample that is nearly an order of magnitude larger than that analyzed by \citet{suzuki2016}. That sample includes several planetary events with mass ratios smaller than $q = 5 \times 10^{-5}$, including OGLE-2019-BLG-1053 \citep[$q=(1.25 \pm 0.13) \times 10^{-5}$;][]{zang2021}, OGLE-2019-BLG-0960 \citep[$q=(1.2\div1.6)\times 10^{-5}$;][]{yee2021}, and KMT-2018-BLG-0029 \citep[$q = (1.81 \pm 0.20)\times 10^{-5}$;][]{gould2020}. Although the statistical analysis of the \textsc{AnomalyFinder} sample is still ongoing and its results were not published at the time of writing this review, we are certain that it will provide strong constraints on the shape of the low-mass planets mass-ratio function.

\section{Galactic distribution of exoplanets}

A sample of planetary microlensing events can be used to check how the frequency of planets changes with the position in the Galaxy. Such an investigation was first conducted by \citet{penny2016} who found that planets in the Galactic bulge seem to be less common than in the Galactic disk. However, this result is very sensitive to a small number of $\pi_\mathrm{E}$ measurements \citep{penny2016}, some of which are susceptible to systematic uncertainties. 
The planet occurrence rate as a function of the Galactocentric distance was studied by \citet{koshimoto2021} but the dependence turned out to be statistically insignificant. One can compare the planet occurrence rate between Galactic disk and bulge, or in more general position in the Galaxy, using microlensing parallaxes measured from the comparison of the ground-based and satellite-based light curves \citep{zhu2017}. Such satellite observations were already collected by the \textit{Spitzer} satellite for a large number of events that were selected using well-defined criteria \citep{yee2015b}. The full analysis of these data has not yet been presented.

\section{Occurrence Rates of Free-floating Planets}

Gravitational microlensing can be also employed to search for and study an elusive population of free-floating planets. These objects, sometimes referred to as rogue planets or unbound planets, are thought to be inevitable outcomes of planet formation processes. There is no universally agreed definition of free-floating planets: For the purpose of this chapter, any object with a mass smaller than the deuterium-burning limit ($\approx 13\,M_{\rm J}$, where $M_{\rm J}$ is the mass of Jupiter) that is not gravitationally bound to a more massive celestial body is called a free-floating planet.

Various processes can lead to the formation of free-floating planets. Such objects may form via gravitational collapse, in a similar way as stars \citep[e.g.,][]{padoan2002,hennebelle2008}, or as failed stellar embryos \citep[e.g.,][]{reipurth2001,whitworth2004}. If they are massive ($\gtrsim 4\div5\,M_{\rm J}$), nearby, and young (at most several Myr) enough, they can be directly detected. Indeed, many planetary-mass objects and objects on a boundary between planets and brown dwarfs were discovered in nearby star-forming regions \citep[e.g.,][]{pena2012,scholz2012}, young associations \citep[e.g.,][]{gagne2017,lodieu2021,miretroig2022}, as well as in the solar neighborhood \citep[e.g.,][]{kirkpatrick2019,kirkpatrick2021}. 

Free-floating planets may also have formed in protoplanetary disks, in a way similar to that in which ordinary extrasolar planets form, either by core accretion \citep{pollack1996} or disk instability \citep{boss1998}, and subsequently ejected from their parent planetary systems. Various processes may be responsible for the ejections. The planet--planet scattering \citep[e.g.,][]{rasio1996,weidenschilling1996,lin1997,chatterjee2008} is the most likely one. 
At least 75\% of systems with giant planets (if not 90--95\%) should have experienced the planet--planet scattering in the past \citep[according to][and references therein]{raymond2022} , in order to explain the observed distribution of eccentricities of giant planets.
These dynamical interactions are likely to disrupt the orbits of inner terrestrial planets or planetary embryos and ultimately lead to their ejection \citep{veras2005,matsumura2013,carrera2016,matsumoto2020}, as demonstrated in extensive simulations by \citet{pfyffer2015}, \citet{ma2016}, or \citet{barclay2017}, among others. Other mechanisms capable of ejecting planets from their parent systems include dynamical interactions in binary and multiple star systems \citep[e.g.,][]{kaib2013,sutherland2016}, interactions in stellar clusters and star-forming regions \citep[e.g.,][]{spurzem2009,parker2012,vanelteren2019,daffern2022,rickman2023}, interactions in gravitationally unstable protoplanetary disks \citep{boss2023}, stellar fly-bys \citep[e.g.,][]{malmberg2011,bailey2019}, or the effects of the post-main-sequence evolution of the host star \citep[e.g.,][]{veras2011}. The ejected planets are expected to be predominantly of low-mass, rendering it virtually impossible to observe them directly.

Fortunately, free-floating planets, even the least massive ones, may be detected by means of their gravity, via gravitational microlensing. There are, however, two major factors that limit their detectability. First, the smaller the mass of the planet, the shorter the duration of the event. The Einstein timescale $t_{\rm E}$ of the microlensing event depends on its angular Einstein radius $\theta_{\rm E}$ and the relative lens--source proper motion $\mu_{\rm rel}$:
\begin{equation}
t_{\rm E} = \frac{\theta_{\rm E}}{\mu_{\rm rel}} = 0.11\,\mathrm{d} \left(\frac{M}{1\,M_{\oplus}}\right)^{1/2} \left(\frac{\pi_{\rm rel}}{0.1\,\rm{mas}}\right)^{1/2} \left(\frac{\mu_{\rm rel}}{5\,\rm{mas\,yr}^{-1}}\right)^{-1},
\end{equation}
where $M$ is the mass of the planet and $\pi_{\rm rel}$ is the relative lens--source parallax. Microlensing events by Jupiter-mass objects have timescales on the order of $1 \div 2$\,d, whereas those by terrestrial-mass planets have timescales of only a few hours. Thus, detecting free-floating planets with microlensing requires high-cadence observations (at least one observation per hour).

Another factor that limits the detectability of free-floating planets is the finite-source effect. This effect is observed when the angular radius of the star being lensed $\theta_*$ becomes comparable to the angular Einstein radius. For typical values of free-floating planets, the Eq.~\ref{eq:rho1} becomes:
\begin{equation}
\rho = 0.4 \left(\frac{\theta_{*}}{0.6\,\mu\rm{as}}\right) \left(\frac{M}{1\,M_{\oplus}}\right)^{-1/2} \left(\frac{\pi_{\rm rel}}{0.1\,\rm{mas}}\right)^{-1/2},
\end{equation}
where $0.6\,\mu\mathrm{as}$ is the angular radius of a $1\,R_{\odot}$ star located in the Galactic center distance. If $\rho \gg 1$, the magnification of the event quickly falls as $A = 1 + 2/\rho^2$. Therefore, the least massive planets produce microlensing signals of a low amplitude, whose light curve shape may resemble that of an ``inverted'' planetary transit. On a positive note, the detection of finite-source effects in the light curve of the event allows us to directly determine the angular Einstein radius of the planet, which further constrains the lens mass.

The first claims about a possible population of free-floating planets did not last long. 
\citet{sahu2001} observed stars behind the globular cluster M22 with the \textit{Hubble Space Telescope} and found six extremely short, unresolved ``events'' (lasting less than 6\,min each).
However, it was quickly realized that these were artifacts caused by cosmic ray hits rather than genuine astrophysical sources \citep{sahu2002}. 

Another attempt to study the population of free-floating planets in the Milky Way was made by \citet{sumi2011}, who analyzed the light curves of 474 microlensing events detected by the MOA survey during the years 2006--2007. The MOA group was the first to conduct a microlensing survey of the Galactic bulge with a high cadence (10$\div$50\,min), hence, MOA observations were sensitive to the shortest-timescale events. \citet{sumi2011} found an apparent excess of short-timescale events ($t_{\rm E}=0.5\div2$\,d), which they claimed was due to a population of about two Jupiter-mass free-floating planets per star. Over a decade later, however, these observations turned out to be severely affected by systematic errors due to differential refraction \citep{koshimoto2023}. Nevertheless, the work by \citet{sumi2011} sparked many theoretical studies on the origin of free-floating planets and opened up the new field.

The claims by \citet{sumi2011} were met with some skepticism in the community, partly because it was difficult to explain how such a large population of Jupiter-mass free-floaters could have formed \citep[e.g.,][]{veras2012}, and partly because the infrared surveys of young clusters and star-forming regions did not find that many planetary-mass objects \citep[e.g.,][]{pena2012,scholz2012}. The start of the fourth phase of the OGLE survey in 2010 \citep{udalski2015b} raised hopes for an independent evaluation of the population of free-floating planets. OGLE observed about 50\,million stars in an area greater than 12\,deg$^2$ at a high cadence ($20 \div 60$\,min).

\begin{figure}
\includegraphics[width=\textwidth]{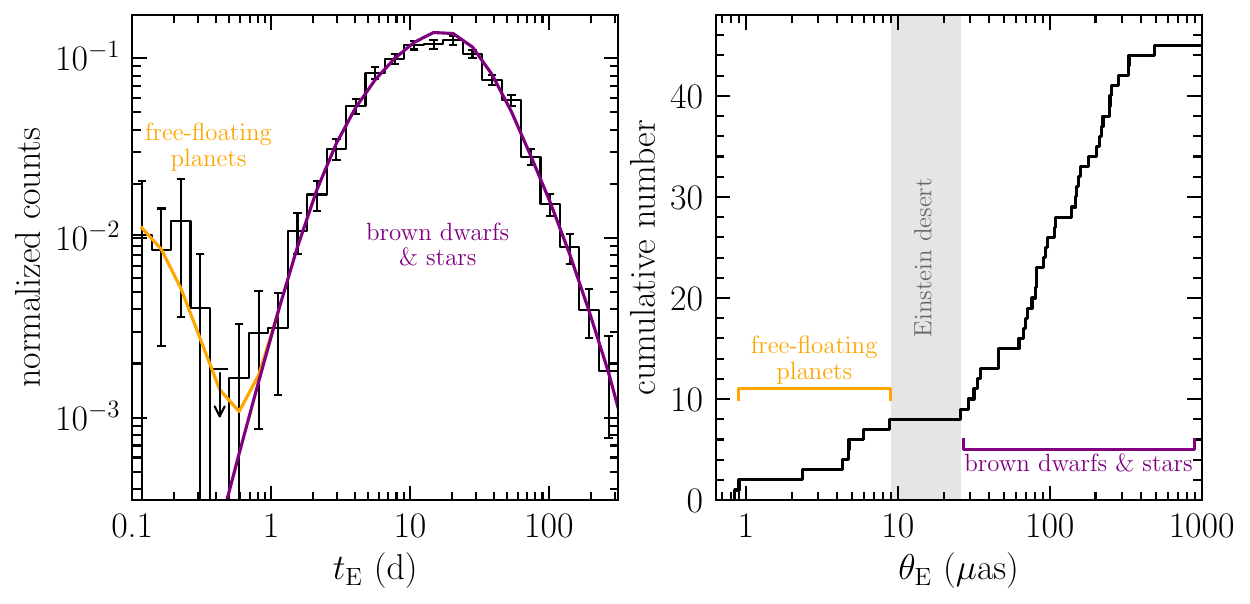}
\caption{Both the distribution of Einstein timescales (left) and angular Einstein radii (right) of single-lens microlensing events indicate the presence of two populations of objects: free-floating planet candidates, and brown dwarfs and stars. These populations are separated by the so-called Einstein desert. Data were taken from \citet{mroz2017, mroz2019, mroz2020a, ryu2021, gould2022, koshimoto2023}.}
\label{fig:ffp}
\end{figure}

The results of a nearly six-year-long observing campaign by the OGLE survey were reported by \citet{mroz2017}, who analyzed the light curves of 2617 microlensing events detected during the years 2010--2015, and measured the distribution of their Einstein timescales. \citet{mroz2017} did not find significant excess of events with timescales $t_{\rm E}=1\div 2$\,d, placing a 95\% upper limit on the frequency of Jupiter-mass free-floating planets of 0.25 planets per main-sequence star. Moreover, they detected six events with extremely short timescales, $t_{\rm E}=0.1 \div0.5$\,d, which were indicative of a population of Earth-mass and super-Earth-mass free-floating planets, as predicted by planet-formation theories. The distribution of Einstein timescales of events detected by \citet{mroz2017} (corrected for selection biases) is shown in the left panel of Fig.~\ref{fig:ffp}.

The discovery of this new population of low-mass free-floating planet candidates coincided with the start of the KMTNet survey \citep{kim2016}. By combining data from OGLE and KMTNet, \citet{mroz2018} detected OGLE-2016-BLG-1540, the first short-timescale ($t_{\rm E} = 0.320 \pm 0.003$\,d) event exhibiting finite-source effects, which enabled the authors to measure its angular Einstein radius. This discovery was soon followed by the detection of a handful of events exhibiting finite-source effects by OGLE, KMTNet, and other surveys \citep{mroz2019,mroz2020a,mroz2020b,ryu2021,kim2021,koshimoto2023}, including likely terrestrial-mass planets \citep{mroz2020a,koshimoto2023}. These detections greatly strengthened the conclusions of \citet{mroz2017}, as they proved the existence of a population of microlensing events with small angular Einstein radii ($\lesssim 10\,\mu\mathrm{as}$) and short timescales ($t_{\rm E} \lesssim 0.5\,\mathrm{d}$).

The measurements of the angular Einstein radii are particularly important because, unlike Einstein timescales, they do not depend on the relative lens--source proper motion. % GG below
In particular, \citet{kim2021} and \citet{ryu2021} noticed that angular Einstein radii of free-floating planet candidates are separated by a gap ($10\,\mu\mathrm{as}\lesssim \theta_{\rm E} \lesssim 30\,\mu\mathrm{as}$), dubbed the ``Einstein desert'', from those of brown dwarfs and stars (see the right panel of Fig.~\ref{fig:ffp}). A similar gap is seen in the distribution of Einstein timescales measured by \citet{mroz2017}.

The Einstein desert was a subject of a detailed investigation by \citet{gould2022}, who systematically analyzed giant-source events exhibiting finite-source effects in the 2016--2019 KMTNet microlensing data set. Thanks to this systematic approach to searches for events with finite-source effects, \citet{gould2022} estimated a simple functional form of the detection efficiency as a function of $\theta_{\rm E}$ (in good agreement with that calculated with extensive simulations carried out by \citet{sumi2023}). \citet{gould2022} demonstrated that the distribution of angular Einstein radii of events observed by KMTNet is consistent with a power law distribution of masses of free-floating planet candidates: $dN/d\log M = (0.4 \pm 0.2) (M/38\,M_{\oplus})^{-p}$/star, with $0.9 \lesssim p \lesssim 1.2$. The power-law index $p < 0.9$ would be inconsistent with the Einstein desert, while $p>1.2$ would be allowed by the data, but it was deemed improbable based on physical arguments.

The power-law mass function of free-floating planet candidates was confirmed by the analysis of a statistical sample of about 3500 microlensing events detected by the MOA survey in 2006--2014 \citep{koshimoto2023,sumi2023}. They were the first to calculate the event detection efficiency as a function of both $t_{\rm E}$ and $\theta_{\rm E}$. A detailed statistical analysis of the MOA data set enabled \citet{sumi2023} to measure the mass function of free-floating planet candidates $dN/d\log M = (0.49^{+0.12}_{-0.32}) (M/38\,M_{\oplus})^{-p}$/star, with $p=0.94^{+0.47}_{-0.27}$, in remarkable agreement with that found by \citet{gould2022}. The \citet{sumi2023} mass function extends down to masses of just $0.33\,M_{\oplus}$, but its shape at such low masses is poorly constrained because there is only a single detection in this mass regime.

A general picture of the population of free-floating planet candidates emerges from the three major studies discussed \citep{mroz2017,gould2022,sumi2023}. These studies are all based on independent data sets and independent data analysis methods, yet they converge to the same conclusion that the mass function of free-floating planet candidates can be approximated by a power law with the index $p \approx 1$ and predict about $\sim 7^{+7}_{-5}$ free-floaters more massive than $1\,M_{\oplus}$ per star. The number of lower-mass objects is even greater, it approximately scales as $(M_{\rm min}/1\,M_{\oplus})^p$, where $M_{\rm min}$ is the minimal mass of free-floating planets considered. \citet{gould2022} considered planets more massive than $\approx 1\,M_{\oplus}$, whereas the study of \citet{sumi2023} was sensitive to even less massive objects ($\gtrsim 0.3\,M_{\oplus}$). The results of all three studies \citep{mroz2017,gould2022,sumi2023} seem to indicate that the number of free-floating planet candidates is of the same order (or even greater) than the number of gravitationally bound planets. If confirmed, this would indicate that ejections may play a major role during the planet formation and shape the mass function of bound exoplanets.

All these studies, however, suffer from the same caveat. Light curves of short-timescale microlensing events detected in the OGLE, KMTNet, and MOA data are consistent with those by isolated objects. However, it cannot be ruled out that all (or some) detected planets may have a widely separated stellar companion. If the host star were located sufficiently far from the lensing object (the planet), it would not leave any detectable signatures in the light curve of the event, rendering it difficult to distinguish between a free-floating and a wide-orbit planet scenario \citep[e.g.,][]{han2003,han2005}. The available photometric data enable us to rule out the presence of a putative host star within only approximately five Einstein radii (projected separation of $\approx 12$\,au) of the planet \citep[e.g.,][]{mroz2018,kim2021}. We note that the widest separation low-mass microlensing planet has $s = 5.25$ \citep{poleski2014c}. It is thus possible that some (or even all) of the detected objects are actually orbiting a distant host star in a wide orbit. For example, the distribution of the smallest $\theta_\mathrm{E}$ events found by \citet{gould2022} was considered consistent with being fully caused by the wide-orbit planet distribution derived by \citet{poleski21}. 
Adaptive optics observations of free-floating planet candidates ruled out some possible host stars \citep{mroz2023}. 
However, deeper imaging observations are needed to distinguish between the free-floating planet and wide-orbit planet hypotheses, such observations will be possible with the \textit{JWST} or the next generation of large 30-m-class ground-based telescopes \citep[e.g.,][]{gould2016}.

There is no doubt that the population of low-mass free-floating planet candidates that were uncovered by ground-based surveys will be subject to more detailed investigations in the future. The planned space-based microlensing experiments are especially promising because they enable almost continuous coverage of light curves, and the exquisite space-based photometry allows one to detect low-amplitude events \citep[e.g.,][]{henderson2016b}. The first attempt to carry out a space-based microlensing survey was the Campaign~9 of the \textit{K2} mission (\textit{Kepler} satellite), which observed an area of $3.7\,\mathrm{deg}^2$ toward the Galactic bulge from 2016 April 22 through July 2 \citep{henderson2016}. These observations were obtained before \citet{mroz2017} results were published, hence, the experiment was planned based on \citet{sumi2011} free-floating planet rate. The \textit{Kepler} spacecraft was not designed to observe dense Galactic bulge fields. Nevertheless, \citet{mcdonald2021} detected four candidates for short-duration microlensing events. Their origin remains unknown, unfortunately, because \citet{mcdonald2021} were able to measure only the effective timescale, $t_{\rm eff} = t_{\rm E} u_0$, which is a combination of the Einstein timescale and the impact parameter of the event.

The \textit{Roman} telescope is planned to carry out a time-domain near-infrared survey of the Galactic bulge in the late 2020s and early 2030s \citep{penny2019}. \textit{Roman} is designed to observe the Galactic bulge during six 72-d-long seasons. Equipped with a 2.4\,m mirror and HAWAII~HgCdTe detectors with a field of view of 0.282\,deg$^2$, \textit{Roman} is predicted to detect $\approx 27,000$ microlensing events in total \citep{penny2019}. 
% GG below
\citet{johnson2020} estimated that \textit{Roman} may detect at least 250 free-floating planets with masses down to that of Mars, assuming their mass function is identical to that of bound planets \citep{cassan12}. Assuming the mass function of \citet{sumi2023}, \textit{Roman} is expected to detect $988^{+1848}_{-566}$ free-floating planets. Similarly, the \textit{Earth 2.0 (ET)} mission by the Chinese Academy of Sciences is a proposed space-based telescope designed to carry out transit and microlensing surveys \citep{ge2022}, following the arguments put forward by \citet{gould2021}. Its planned launch date is 2027. The \textit{ET} mission will consist of seven 30-cm-class telescopes, one of which is planned to be dedicated to microlensing observations. The microlensing telescope is designed to have the same field of view (4\,deg$^2$) and pixel scale ($0.4''$) as the KMTNet telescopes. The \textit{ET} mission is predicted to detect about 600 free-floating planets assuming the mass function of \citet{gould2022} and the duration of the survey of 4\,yr. 

The planned space-based missions make also promise for direct measurements of the masses of free-floating planet candidates, which is possible owing to the satellite microlensing parallax effects (the same event simultaneously observed from the ground and a sufficiently separated space-based observatory should look differently). Events detected by the \textit{Roman} or the \textit{Earth 2.0} missions may be additionally observed with ground-based \citep{zhu2016,street2018,gould2021} and space-based facilities: \textit{Euclid} \citep{bachelet2019,ban2020,bachelet2022}, \textit{Chinese Space Station Telescope} \citep{yan2022}, or \textit{CLEoPATRA} \citep{barry2022}. These additional observations will enable us to better understand the population of free-floating planets and the processes that occur at the early stages of planet formation.

\section{Cross-References}
\begin{itemize}
\item{Finding Planets via Gravitational Microlensing}
% https://link.springer.com/referenceworkentry/10.1007/978-3-319-55333-7_120
\item{Microlensing Surveys for Exoplanet Research (OGLE Survey Perspective)}
% https://link.springer.com/referenceworkentry/10.1007/978-3-319-55333-7_123
\item{Microlensing Surveys for Exoplanet Research (MOA)}
% https://link.springer.com/referenceworkentry/10.1007/978-3-319-55333-7_122
\item{Korea Microlensing Telescope Network}
% https://link.springer.com/referenceworkentry/10.1007/978-3-319-55333-7_124
\end{itemize}

\begin{acknowledgement}
We thank Andrzej Udalski, Weicheng Zang, and Aparna Bhattacharya for their comments on the early draft of the manuscript.
This research has made use of the NASA Exoplanet Archive, which is operated by the California Institute of Technology, under contract with the National Aeronautics and Space Administration under the Exoplanet Exploration Program. 
Work by RP was supported by the Polish National Agency for Academic Exchange grant ``Polish Returns 2019.'' \end{acknowledgement}

\bibliographystyle{spbasicHBexo}  %for bibtex
\bibliography{pap} %for bibtex-example

\begin{thebibliography}{133}
\providecommand{\natexlab}[1]{#1}
\providecommand{\url}[1]{{#1}}
\providecommand{\urlprefix}{URL }
\expandafter\ifx\csname urlstyle\endcsname\relax
  \providecommand{\doi}[1]{DOI~\discretionary{}{}{}#1}\else
  \providecommand{\doi}{DOI~\discretionary{}{}{}\begingroup
  \urlstyle{rm}\Url}\fi
\providecommand{\eprint}[2][]{\url{#2}}

\bibitem[{{Alcock} et~al.(2001){Alcock}, {Allsman}, {Alves}, {Axelrod},
  {Becker}, {Bennett}, {Cook}, {Drake}, {Freeman}, {Geha}, {Griest}, {Keller},
  {Lehner}, {Marshall}, {Minniti}, {Nelson}, {Peterson}, {Popowski}, {Pratt},
  {Quinn}, {Stubbs}, {Sutherland}, {Tomaney}, {Vandehei}, and
  {Welch}}]{alcock2001}
{Alcock} C, {Allsman} RA, {Alves} DR et~al. (2001) {Direct detection of a
  microlens in the Milky Way}. \nat 414(6864):617--619

\bibitem[{{Bachelet} and {Penny}(2019)}]{bachelet2019}
{Bachelet} E {Penny} M (2019) {WFIRST and EUCLID: Enabling the Microlensing
  Parallax Measurement from Space}. \apjl 880(2):L32

\bibitem[{{Bachelet} et~al.(2022){Bachelet}, {Specht}, {Penny}, {Hundertmark},
  {Awiphan}, {Beaulieu}, {Dominik}, {Kerins}, {Maoz}, {Meade}, {Nucita},
  {Poleski}, {Ranc}, {Rhodes}, and {Robin}}]{bachelet2022}
{Bachelet} E, {Specht} D, {Penny} M et~al. (2022) {Euclid-Roman joint
  microlensing survey: Early mass measurement, free floating planets, and
  exomoons}. \aap 664:A136

\bibitem[{{Bailey} and {Fabrycky}(2019)}]{bailey2019}
{Bailey} N {Fabrycky} D (2019) {Stellar Flybys Interrupting Planet-Planet
  Scattering Generates Oort Planets}. \aj 158(2):94

\bibitem[{{Ban}(2020)}]{ban2020}
{Ban} M (2020) {Probability of simultaneous parallax detection for
  free-floating planet microlensing events near Galactic Centre}. \mnras
  494(3):3235--3252

\bibitem[{{Barclay} et~al.(2017){Barclay}, {Quintana}, {Raymond}, and
  {Penny}}]{barclay2017}
{Barclay} T, {Quintana} EV, {Raymond} SN {Penny} MT (2017) {The Demographics of
  Rocky Free-floating Planets and their Detectability by WFIRST}. \apj
  841(2):86

\bibitem[{{Barry} et~al.(2022){Barry}, {Olmschenk}, {Ravizza}, {Bond},
  {Cervantes}, {Ishitani-Silva}, {Korde-patel}, {Rattenbury}, {Rau}, {Sumi},
  and {Wyrwas}}]{barry2022}
{Barry} RK, {Olmschenk} G, {Ravizza} F et~al. (2022) {CLEoPATRA:
  contemporaneous lensing parallax and autonomous transient assay}. In: {Coyle}
  LE, {Matsuura} S {Perrin} MD (eds) Space Telescopes and Instrumentation 2022:
  Optical, Infrared, and Millimeter Wave, Society of Photo-Optical
  Instrumentation Engineers (SPIE) Conference Series, vol 12180, p 121800D,
  \doi{10.1117/12.2620310}

\bibitem[{Batista(2018)}]{batista2018}
Batista V (2018) Finding planets via gravitational microlensing. In: Deeg HJ
  Belmonte JA (eds) Handbook of Exoplanets, Springer International Publishing,
  Cham, pp 659--687, \doi{10.1007/978-3-319-55333-7_120}

\bibitem[{{Batista} et~al.(2015){Batista}, {Beaulieu}, {Bennett}, {Gould},
  {Marquette}, {Fukui}, and {Bhattacharya}}]{batista2015}
{Batista} V, {Beaulieu} JP, {Bennett} DP et~al. (2015) {Confirmation of the
  OGLE-2005-BLG-169 Planet Signature and Its Characteristics with Lens-Source
  Proper Motion Detection}. \apj 808(2):170

\bibitem[{{Bennett} et~al.(2015){Bennett}, {Bhattacharya}, {Anderson}, {Bond},
  {Anderson}, {Barry}, {Batista}, {Beaulieu}, {DePoy}, {Dong}, {Gaudi},
  {Gilbert}, {Gould}, {Pfeifle}, {Pogge}, {Suzuki}, {Terry}, and
  {Udalski}}]{bennett2015}
{Bennett} DP, {Bhattacharya} A, {Anderson} J et~al. (2015) {Confirmation of the
  Planetary Microlensing Signal and Star and Planet Mass Determinations for
  Event OGLE-2005-BLG-169}. \apj 808(2):169

\bibitem[{{Bhattacharya} et~al.(2018){Bhattacharya}, {Beaulieu}, {Bennett},
  {Anderson}, {Koshimoto}, {Lu}, {Batista}, {Blackman}, {Bond}, {Fukui},
  {Henderson}, {Hirao}, {Marquette}, {Mroz}, {Ranc}, and
  {Udalski}}]{bhattacharya2018}
{Bhattacharya} A, {Beaulieu} JP, {Bennett} DP et~al. (2018) {WFIRST Exoplanet
  Mass-measurement Method Finds a Planetary Mass of 39 {\ensuremath{\pm}} 8 M
  $_{{\ensuremath{\oplus}}}$ for OGLE-2012-BLG-0950Lb}. \aj 156(6):289

\bibitem[{{Boss}(1998)}]{boss1998}
{Boss} AP (1998) {Evolution of the Solar Nebula. IV. Giant Gaseous Protoplanet
  Formation}. \apj 503(2):923--937

\bibitem[{{Boss}(2023)}]{boss2023}
{Boss} AP (2023) {Orbital Migration of Protoplanets in a Marginally
  Gravitationally Unstable Disk. II. Migration, Merging, and Ejection}. \apj
  943(2):101

\bibitem[{{Carrera} et~al.(2016){Carrera}, {Davies}, and
  {Johansen}}]{carrera2016}
{Carrera} D, {Davies} MB {Johansen} A (2016) {Survival of habitable planets in
  unstable planetary systems}. \mnras 463(3):3226--3238

\bibitem[{{Cassan} et~al.(2012){Cassan}, {Kubas}, {Beaulieu}, {Dominik},
  {Horne}, {Greenhill}, {Wambsganss}, {Menzies}, {Williams}, {J{\o}rgensen},
  {Udalski}, {Bennett}, {Albrow}, {Batista}, {Brillant}, {Caldwell}, {Cole},
  {Coutures}, {Cook}, {Dieters}, {Dominis Prester}, {Donatowicz}, {Fouqu{\'e}},
  {Hill}, {Kains}, {Kane}, {Marquette}, {Martin}, {Pollard}, {Sahu}, {Vinter},
  {Warren}, {Watson}, {Zub}, {Sumi}, {Szyma{\'n}ski}, {Kubiak}, {Poleski},
  {Soszynski}, {Ulaczyk}, {Pietrzy{\'n}ski}, and {Wyrzykowski}}]{cassan12}
{Cassan} A, {Kubas} D, {Beaulieu} JP et~al. (2012) {One or more bound planets
  per Milky Way star from microlensing observations}. \nat 481(7380):167--169

\bibitem[{{Chatterjee} et~al.(2008){Chatterjee}, {Ford}, {Matsumura}, and
  {Rasio}}]{chatterjee2008}
{Chatterjee} S, {Ford} EB, {Matsumura} S {Rasio} FA (2008) {Dynamical Outcomes
  of Planet-Planet Scattering}. \apj 686(1):580--602

\bibitem[{{Clanton} and {Gaudi}(2014{\natexlab{a}})}]{clanton2014a}
{Clanton} C {Gaudi} BS (2014{\natexlab{a}}) {Synthesizing Exoplanet
  Demographics from Radial Velocity and Microlensing Surveys. I. Methodology}.
  \apj 791(2):90

\bibitem[{{Clanton} and {Gaudi}(2014{\natexlab{b}})}]{clanton2014b}
{Clanton} C {Gaudi} BS (2014{\natexlab{b}}) {Synthesizing Exoplanet
  Demographics from Radial Velocity and Microlensing Surveys. II. The Frequency
  of Planets Orbiting M Dwarfs}. \apj 791(2):91

\bibitem[{{Clanton} and {Gaudi}(2016)}]{clanton2016}
{Clanton} C {Gaudi} BS (2016) {Synthesizing Exoplanet Demographics: A Single
  Population of Long-period Planetary Companions to M Dwarfs Consistent with
  Microlensing, Radial Velocity, and Direct Imaging Surveys}. \apj 819(2):125

\bibitem[{{Daffern-Powell} et~al.(2022){Daffern-Powell}, {Parker}, and
  {Quanz}}]{daffern2022}
{Daffern-Powell} EC, {Parker} RJ {Quanz} SP (2022) {The Great Planetary Heist:
  theft and capture in star-forming regions}. \mnras 514(1):920--934

\bibitem[{{Fukui} et~al.(2019){Fukui}, {Suzuki}, {Koshimoto}, {Bachelet},
  {Vanmunster}, {Storey}, {Maehara}, {Yanagisawa}, {Yamada}, {Yonehara},
  {Hirano}, {Bennett}, {Bozza}, {Mawet}, {Penny}, {Awiphan}, {Oksanen},
  {Heintz}, {Oberst}, {B{\'e}jar}, {Casasayas-Barris}, {Chen}, {Crouzet},
  {Hidalgo}, {Klagyivik}, {Murgas}, {Narita}, {Palle}, {Parviainen},
  {Watanabe}, {Kusakabe}, {Mori}, {Terada}, {de Leon}, {Hernandez}, {Luque},
  {Monelli}, {Monta{\~n}es-Rodriguez}, {Prieto-Arranz}, {Murata}, {Shugarov},
  {Kubota}, {Otsuki}, {Shionoya}, {Nishiumi}, {Nishide}, {Fukagawa}, {Onodera},
  {Villanueva}, {Street}, {Tsapras}, {Hundertmark}, {Kuzuhara}, {Fujita},
  {Beichman}, {Beaulieu}, {Alonso}, {Reichart}, {Kawai}, and
  {Tamura}}]{fukui2019}
{Fukui} A, {Suzuki} D, {Koshimoto} N et~al. (2019) {Kojima-1Lb Is a Mildly Cold
  Neptune around the Brightest Microlensing Host Star}. \aj 158(5):206

\bibitem[{{Gagn{\'e}} et~al.(2017){Gagn{\'e}}, {Faherty}, {Mamajek}, {Malo},
  {Doyon}, {Filippazzo}, {Weinberger}, {Donaldson}, {L{\'e}pine},
  {Lafreni{\`e}re}, {Artigau}, {Burgasser}, {Looper}, {Boucher}, {Beletsky},
  {Camnasio}, {Brunette}, and {Arboit}}]{gagne2017}
{Gagn{\'e}} J, {Faherty} JK, {Mamajek} EE et~al. (2017) {BANYAN. IX. The
  Initial Mass Function and Planetary-mass Object Space Density of the TW HYA
  Association}. \apjs 228(2):18

\bibitem[{{Gaudi}(2012)}]{gaudi2012}
{Gaudi} BS (2012) {Microlensing Surveys for Exoplanets}. \araa 50:411--453

\bibitem[{{Gaudi} and {Sackett}(2000)}]{gaudi00}
{Gaudi} BS {Sackett} PD (2000) {Detection Efficiencies of Microlensing Data
  Sets to Stellar and Planetary Companions}. \apj 528(1):56--73

\bibitem[{{Gaudi} et~al.(2002){Gaudi}, {Albrow}, {An}, {Beaulieu}, {Caldwell},
  {DePoy}, {Dominik}, {Gould}, {Greenhill}, {Hill}, {Kane}, {Martin},
  {Menzies}, {Naber}, {Pel}, {Pogge}, {Pollard}, {Sackett}, {Sahu}, {Vermaak},
  {Vreeswijk}, {Watson}, and {Williams}}]{gaudi2002}
{Gaudi} BS, {Albrow} MD, {An} J et~al. (2002) {Microlensing Constraints on the
  Frequency of Jupiter-Mass Companions: Analysis of 5 Years of PLANET
  Photometry}. \apj 566(1):463--499

\bibitem[{{Ge} et~al.(2022){Ge}, {Zhang}, {Zang}, {Deng}, {Mao}, {Xie}, {Liu},
  {Zhou}, {Willis}, {Huang}, {Howell}, {Feng}, {Zhu}, {Yao}, {Liu}, {Aizawa},
  {Zhu}, {Li}, {Ma}, {Ye}, {Yu}, {Xiang}, {Yu}, {Liu}, {Yang}, {Wang}, {Shi},
  {Fang}, {Zong}, {Liu}, {Zhang}, {Zhang}, {El-Badry}, {Shen}, {Tam}, {Hu},
  {Yang}, {Zou}, {Wu}, {Lei}, {Wei}, {Wu}, {Sun}, {Wang}, {Zhang}, {Xu},
  {Yang}, {Li}, {Xiang}, {Wang}, {Wang}, {Zhang}, {Jia}, {Yuan}, {Zhang},
  {Xuesong Wang}, {Gan}, {Wang}, {Zhao}, {Liu}, {Wei}, {Kang}, {Yang}, {Qi},
  {Liu}, {Zhang}, {Zhu}, {Zhou}, {Zhang}, {Yu}, {Zhang}, {Li}, {Tang}, {Wang},
  {Wang}, {Li}, {Cheng}, {Shen}, {Li}, {Pan}, {Yang}, {Gao}, {Song}, {Wang},
  {Zhang}, {Chen}, {Wang}, {Zhang}, {Wang}, {Zeng}, {Zheng}, {Zhu}, {Guo},
  {Zhang}, {Li}, {Wen}, {Feng}, {Chen}, {Chen}, {Han}, {Yang}, {Wang}, {Duan},
  {Huang}, {Liang}, {Bi}, {Gai}, {Ge}, {Guo}, {Huang}, {Li}, {Li}, {Li},
  {Yuxi}, {Lu}, {Rix}, {Shi}, {Song}, {Tang}, {Ting}, {Wu}, {Wu}, {Yang},
  {Yin}, {Gould}, {Lee}, {Dong}, {Yee}, {Shvartzvald}, {Yang}, {Kuang},
  {Zhang}, {Liao}, {Qi}, {Yang}, {Zhang}, {Jiang}, {Ou}, {Li}, {Beck},
  {Bedding}, {Campante}, {Chaplin}, {Christensen-Dalsgaard}, {Garc{\'\i}a},
  {Gaulme}, {Gizon}, {Hekker}, {Huber}, {Khanna}, {Li}, {Mathur}, {Miglio},
  {Mosser}, {Ong}, {Santos}, {Stello}, {Bowman}, {Lares-Martiz}, {Murphy},
  {Niu}, {Ma}, {Moln{\'a}r}, {Fu}, {De Cat}, {Su}, and {consortium}}]{ge2022}
{Ge} J, {Zhang} H, {Zang} W et~al. (2022) {ET White Paper: To Find the First
  Earth 2.0}. arXiv e-prints arXiv:2206.06693

\bibitem[{{Gould}(1992)}]{gould1992b}
{Gould} A (1992) {Extending the MACHO Search to approximately 10 6 M sub sun}.
  \apj 392:442

\bibitem[{{Gould}(1994)}]{gould1994}
{Gould} A (1994) {Proper Motions of MACHOs}. \apjl 421:L71

\bibitem[{{Gould}(2016{\natexlab{a}})}]{gould2016}
{Gould} A (2016{\natexlab{a}}) {Microlensing by Kuiper, Oort, and Free-Floating
  Planets}. Journal of Korean Astronomical Society 49(4):123--126

\bibitem[{{Gould}(2016{\natexlab{b}})}]{gould2016rev}
{Gould} A (2016{\natexlab{b}}) {Microlensing Planets}. In: {Bozza} V, {Mancini}
  L {Sozzetti} A (eds) Methods of Detecting Exoplanets: 1st Advanced School on
  Exoplanetary Science, Astrophysics and Space Science Library, vol 428, p 135,
  \doi{10.1007/978-3-319-27458-4_3}

\bibitem[{{Gould} and {Loeb}(1992)}]{gould1992}
{Gould} A {Loeb} A (1992) {Discovering Planetary Systems through Gravitational
  Microlenses}. \apj 396:104

\bibitem[{{Gould} et~al.(2010){Gould}, {Dong}, {Gaudi}, {Udalski}, {Bond},
  {Greenhill}, {Street}, {Dominik}, {Sumi}, {Szyma{\'n}ski}, {Han}, {Allen},
  {Bolt}, {Bos}, {Christie}, {DePoy}, {Drummond}, {Eastman}, {Gal-Yam},
  {Higgins}, {Janczak}, {Kaspi}, {Koz{\l}owski}, {Lee}, {Mallia}, {Maury},
  {Maoz}, {McCormick}, {Monard}, {Moorhouse}, {Morgan}, {Natusch}, {Ofek},
  {Park}, {Pogge}, {Polishook}, {Santallo}, {Shporer}, {Spector}, {Thornley},
  {Yee}, {{\ensuremath{\mu}}FUN Collaboration}, {Kubiak}, {Pietrzy{\'n}ski},
  {Soszy{\'n}ski}, {Szewczyk}, {Wyrzykowski}, {Ulaczyk}, {Poleski}, {OGLE
  Collaboration}, {Abe}, {Bennett}, {Botzler}, {Douchin}, {Freeman}, {Fukui},
  {Furusawa}, {Hearnshaw}, {Hosaka}, {Itow}, {Kamiya}, {Kilmartin}, {Korpela},
  {Lin}, {Ling}, {Makita}, {Masuda}, {Matsubara}, {Miyake}, {Muraki}, {Nagaya},
  {Nishimoto}, {Ohnishi}, {Okumura}, {Perrott}, {Philpott}, {Rattenbury},
  {Saito}, {Sako}, {Sullivan}, {Sweatman}, {Tristram}, {von Seggern}, {Yock},
  {MOA Collaboration}, {Albrow}, {Batista}, {Beaulieu}, {Brillant}, {Caldwell},
  {Calitz}, {Cassan}, {Cole}, {Cook}, {Coutures}, {Dieters}, {Dominis Prester},
  {Donatowicz}, {Fouqu{\'e}}, {Hill}, {Hoffman}, {Jablonski}, {Kane}, {Kains},
  {Kubas}, {Marquette}, {Martin}, {Martioli}, {Meintjes}, {Menzies},
  {Pedretti}, {Pollard}, {Sahu}, {Vinter}, {Wambsganss}, {Watson}, {Williams},
  {Zub}, {PLANET Collaboration}, {Allan}, {Bode}, {Bramich}, {Burgdorf},
  {Clay}, {Fraser}, {Hawkins}, {Horne}, {Kerins}, {Lister}, {Mottram},
  {Saunders}, {Snodgrass}, {Steele}, {Tsapras}, {RoboNet Collaboration},
  {J{\o}rgensen}, {Anguita}, {Bozza}, {Calchi Novati}, {Harps{\o}e}, {Hinse},
  {Hundertmark}, {Kj{\ae}rgaard}, {Liebig}, {Mancini}, {Masi}, {Mathiasen},
  {Rahvar}, {Ricci}, {Scarpetta}, {Southworth}, {Surdej}, {Th{\"o}ne}, and
  {MiNDSTEp Consortium}}]{gould10}
{Gould} A, {Dong} S, {Gaudi} BS et~al. (2010) {Frequency of Solar-like Systems
  and of Ice and Gas Giants Beyond the Snow Line from High-magnification
  Microlensing Events in 2005-2008}. \apj 720(2):1073--1089

\bibitem[{{Gould} et~al.(2020){Gould}, {Ryu}, {Calchi Novati}, {Zang},
  {Albrow}, {Chung}, {Han}, {Hwang}, {Jung}, {Shin}, {Shvartzvald}, {Yee},
  {Cha}, {Kim}, {Kim}, {Kim}, {Lee}, {Lee}, {Lee}, {Park}, {Pogge}, {Beichman},
  {Bryden}, {Carey}, {Gaudi}, {Henderson}, {Zhu}, {Fouque}, {Penny}, {Petric},
  {Burdullis}, and {Mao}}]{gould2020}
{Gould} A, {Ryu} YH, {Calchi Novati} S et~al. (2020) {KMT-2018-BLG-0029Lb: A
  Very Low Mass-Ratio Spitzer Microlens Planet}. Journal of Korean Astronomical
  Society 53:9--26

\bibitem[{{Gould} et~al.(2021){Gould}, {Zang}, {Mao}, and {Dong}}]{gould2021}
{Gould} A, {Zang} WC, {Mao} S {Dong} SB (2021) {Masses for free-floating
  planets and dwarf planets}. Research in Astronomy and Astrophysics 21(6):133

\bibitem[{{Gould} et~al.(2022{\natexlab{a}}){Gould}, {Han}, {Zang}, {Yang},
  {Hwang}, {Udalski}, {Bond}, {Albrow}, {Chung}, {Jung}, {Ryu}, {Shin},
  {Shvartzvald}, {Yee}, {Cha}, {Kim}, {Kim}, {Kim}, {Lee}, {Lee}, {Lee},
  {Park}, {Pogge}, {KMTNet Collaboration}, {Mr{\'o}z}, {Szyma{\'n}ski},
  {Skowron}, {Poleski}, {Soszy{\'n}ski}, {Pietrukowicz}, {Koz{\l}owski},
  {Ulaczyk}, {Rybicki}, {Iwanek}, {Wrona}, {OGLE Collaboration}, {Abe},
  {Barry}, {Bennett}, {Bhattacharya}, {Fujii}, {Fukui}, {Hirao}, {Silva},
  {Kirikawa}, {Kondo}, {Koshimoto}, {Matsubara}, {Matsumoto}, {Miyazaki},
  {Muraki}, {Okamura}, {Olmschenk}, {Ranc}, {Rattenbury}, {Satoh}, {Sumi},
  {Suzuki}, {Toda}, {Tristram}, {Vandorou}, {Yama}, {Moa Collaboration},
  {Beichman}, {Bryden}, {Novati}, {Gaudi}, {Henderson}, {Penny}, {Jacklin},
  {Stassun}, and {Ukirt Microlensing Team}}]{gould2022a}
{Gould} A, {Han} C, {Zang} W et~al. (2022{\natexlab{a}}) {Systematic KMTNet
  planetary anomaly search. V. Complete sample of 2018 prime-field}. \aap
  664:A13

\bibitem[{{Gould} et~al.(2022{\natexlab{b}}){Gould}, {Jung}, {Hwang}, {Dong},
  {Albrow}, {Chung}, {Han}, {Ryu}, {Shin}, {Shvartzvald}, {Yang}, {Yee},
  {Zang}, {Cha}, {Kim}, {Kim}, {Lee}, {Lee}, {Lee}, {Park}, and
  {Pogge}}]{gould2022}
{Gould} A, {Jung} YK, {Hwang} KH et~al. (2022{\natexlab{b}}) {Free-Floating
  Planets, the Einstein Desert, and 'OUMUAMUA}. Journal of Korean Astronomical
  Society 55:173--194

\bibitem[{{Griest} and {Safizadeh}(1998)}]{griest98}
{Griest} K {Safizadeh} N (1998) {The Use of High-Magnification Microlensing
  Events in Discovering Extrasolar Planets}. \apj 500(1):37--50

\bibitem[{{Han} and {Kang}(2003)}]{han2003}
{Han} C {Kang} YW (2003) {Probing the Spatial Distribution of Extrasolar
  Planets with Gravitational Microlensing}. \apj 596(2):1320--1326

\bibitem[{{Han} et~al.(2005){Han}, {Gaudi}, {An}, and {Gould}}]{han2005}
{Han} C, {Gaudi} BS, {An} JH {Gould} A (2005) {Microlensing Detection and
  Characterization of Wide-Separation Planets}. \apj 618(2):962--972

\bibitem[{{Henderson} and {Shvartzvald}(2016)}]{henderson2016b}
{Henderson} CB {Shvartzvald} Y (2016) {On the Feasibility of Characterizing
  Free-floating Planets with Current and Future Space-based Microlensing
  Surveys}. \aj 152(4):96

\bibitem[{{Henderson} et~al.(2016){Henderson}, {Poleski}, {Penny}, {Street},
  {Bennett}, {Hogg}, {Gaudi}, {K2 Campaign 9 Microlensing Science Team}, {Zhu},
  {Barclay}, {Barentsen}, {Howell}, {Mullally}, {Udalski}, {Szyma{\'n}ski},
  {Skowron}, {Mr{\'o}z}, {Koz{\l}owski}, {Wyrzykowski}, {Pietrukowicz},
  {Soszy{\'n}ski}, {Ulaczyk}, {Pawlak}, {OGLE Project}, {Sumi}, {Abe},
  {Asakura}, {Barry}, {Bhattacharya}, {Bond}, {Donachie}, {Freeman}, {Fukui},
  {Hirao}, {Itow}, {Koshimoto}, {Li}, {Ling}, {Masuda}, {Matsubara}, {Muraki},
  {Nagakane}, {Ohnishi}, {Oyokawa}, {Rattenbury}, {Saito}, {Sharan},
  {Sullivan}, {Tristram}, {Yonehara}, {MOA Collaboration}, {Bachelet},
  {Bramich}, {Cassan}, {Dominik}, {Figuera Jaimes}, {Horne}, {Hundertmark},
  {Mao}, {Ranc}, {Schmidt}, {Snodgrass}, {Steele}, {Tsapras}, {Wambsganss},
  {RoboNet Project}, {Burgdorf}, {J{\o}rgensen}, {Calchi Novati}, {Ciceri},
  {D'Ago}, {Evans}, {Hessman}, {Hinse}, {Husser}, {Mancini}, {Popovas},
  {Rabus}, {Rahvar}, {Scarpetta}, {Skottfelt}, {Southworth}, {Unda-Sanzana},
  {MiNDSTEp Team}, {Bryson}, {Caldwell}, {Haas}, {Larson}, {McCalmont},
  {Packard}, {Peterson}, {Putnam}, {Reedy}, {Ross}, {Van Cleve}, {K2C9
  Engineering Team}, {Akeson}, {Batista}, {Beaulieu}, {Beichman}, {Bryden},
  {Ciardi}, {Cole}, {Coutures}, {Foreman-Mackey}, {Fouqu{\'e}}, {Friedmann},
  {Gelino}, {Kaspi}, {Kerins}, {Korhonen}, {Lang}, {Lee}, {Lineweaver}, {Maoz},
  {Marquette}, {Mogavero}, {Morales}, {Nataf}, {Pogge}, {Santerne},
  {Shvartzvald}, {Suzuki}, {Tamura}, {Tisserand}, and {Wang}}]{henderson2016}
{Henderson} CB, {Poleski} R, {Penny} M et~al. (2016) {Campaign 9 of the K2
  Mission: Observational Parameters, Scientific Drivers, and Community
  Involvement for a Simultaneous Space- and Ground-based Microlensing Survey}.
  \pasp 128(970):124,401

\bibitem[{{Hennebelle} and {Chabrier}(2008)}]{hennebelle2008}
{Hennebelle} P {Chabrier} G (2008) {Analytical Theory for the Initial Mass
  Function: CO Clumps and Prestellar Cores}. \apj 684(1):395--410

\bibitem[{{Hwang} et~al.(2022){Hwang}, {Zang}, {Gould}, {Udalski}, {Bond},
  {Yang}, {Mao}, {Mao}, {Albrow}, {Chung}, {Han}, {Kil Jung}, {Ryu}, {Shin},
  {Shvartzvald}, {Yee}, {Cha}, {Kim}, {Kim}, {Kim}, {Lee}, {Lee}, {Lee},
  {Park}, {Pogge}, {Pogge}, {Mr{\'o}z}, {Poleski}, {Skowron}, {Szyma{\'n}ski},
  {Soszy{\'n}ski}, {Pietrukowicz}, {Koz{\l}owski}, {Ulaczyk}, {Rybicki},
  {Iwanek}, {Wrona}, {Gromadzki}, {Gromadzki}, {Abe}, {Barry}, {Bennett},
  {Bhattacharya}, {Fujii}, {Fukui}, {Hirao}, {Itow}, {Kirikawa}, {Kondo},
  {Koshimoto}, {Munford}, {Matsubara}, {Miyazaki}, {Muraki}, {Olmschenk},
  {Ranc}, {Rattenbury}, {Satoh}, {Shoji}, {Ishitani Silva}, {Sumi}, {Suzuki},
  {Tristram}, {Yonehara}, {Yonehara}, {Zhang}, {Zhu}, {Penny}, {Fouqu{\'e}},
  and {Fouqu{\'e}}}]{hwang2022a}
{Hwang} KH, {Zang} W, {Gould} A et~al. (2022) {Systematic KMTNet Planetary
  Anomaly Search. II. Six New $q < 2 \times 10^{-4}$ Mass-ratio Planets}. \aj
  163(2):43

\bibitem[{{Ida} and {Lin}(2004)}]{ida2004}
{Ida} S {Lin} DNC (2004) {Toward a Deterministic Model of Planetary Formation.
  I. A Desert in the Mass and Semimajor Axis Distributions of Extrasolar
  Planets}. \apj 604(1):388--413

\bibitem[{{Johnson} et~al.(2020){Johnson}, {Penny}, {Gaudi}, {Kerins},
  {Rattenbury}, {Robin}, {Calchi Novati}, and {Henderson}}]{johnson2020}
{Johnson} SA, {Penny} M, {Gaudi} BS et~al. (2020) {Predictions of the Nancy
  Grace Roman Space Telescope Galactic Exoplanet Survey. II. Free-floating
  Planet Detection Rates}. \aj 160(3):123

\bibitem[{{Jung} et~al.(2019){Jung}, {Gould}, {Zang}, {Hwang}, {Ryu}, {Han},
  {Yee}, {Albrow}, {Chung}, {Shin}, {Shvartzvald}, {Zhu}, {Cha}, {Kim}, {Kim},
  {Kim}, {Lee}, {Lee}, {Lee}, {Park}, {Pogge}, {KMTNet Collaboration}, {Penny},
  {Mao}, {Fouqu{\'e}}, {Wang}, and {CFHT Collaboration}}]{jung19}
{Jung} YK, {Gould} A, {Zang} W et~al. (2019) {KMT-2017-BLG-0165Lb: A
  Super-Neptune-mass Planet Orbiting a Sun-like Host Star}. \aj 157(2):72

\bibitem[{{Jung} et~al.(2022){Jung}, {Zang}, {Han}, {Gould}, {Udalski},
  {Albrow}, {Chung}, {Hwang}, {Ryu}, {Shin}, {Shvartzvald}, {Yang}, {Yee},
  {Cha}, {Kim}, {Kim}, {Lee}, {Lee}, {Lee}, {Park}, {Pogge}, {KMTNet
  Collaboration}, {Mr{\'o}z}, {Szyma{\'n}ski}, {Skowron}, {Poleski},
  {Soszy{\'n}ski}, {Pietrukowicz}, {Koz{\l}owski}, {Ulaczyk}, {Rybicki},
  {Iwanek}, {Wrona}, and {OGLE Collaboration}}]{jung2022a}
{Jung} YK, {Zang} W, {Han} C et~al. (2022) {Systematic KMTNet Planetary Anomaly
  Search. VI. Complete Sample of 2018 Sub-prime-field Planets}. \aj 164(6):262

\bibitem[{{Jung} et~al.(2023){Jung}, {Zang}, {Wang}, {Han}, {Gould}, {Udalski},
  {Albrow}, {Chung}, {Hwang}, {Ryu}, {Shin}, {Shvartzvald}, {Yang}, {Yee},
  {Cha}, {Kim}, {Kim}, {Lee}, {Lee}, {Lee}, {Park}, {Pogge}, {KMTNet
  Collaboration}, {Szyma{\'n}ski}, {Skowron}, {Poleski}, {Soszy{\'n}ski},
  {Pietrukowicz}, {Koz{\l}owski}, {Ulaczyk}, {Rybicki}, {Iwanek}, {Wrona},
  {OGLE Collaboration}, {Green}, {Hennerley}, {Marmont}, {Mao}, {Maoz},
  {McCormick}, {Natusch}, {Penny}, {Porritt}, {Zhu}, {Tsinghua Team}, and {FUN
  Follow-Up Team}}]{jung2023a}
{Jung} YK, {Zang} W, {Wang} H et~al. (2023) {Systematic KMTNet Planetary
  Anomaly Search. VIII. Complete Sample of 2019 Subprime Field Planets}. \aj
  165(6):226

\bibitem[{{Kaib} et~al.(2013){Kaib}, {Raymond}, and {Duncan}}]{kaib2013}
{Kaib} NA, {Raymond} SN {Duncan} M (2013) {Planetary system disruption by
  Galactic perturbations to wide binary stars}. \nat 493(7432):381--384

\bibitem[{{Kim} et~al.(2021){Kim}, {Hwang}, {Gould}, {Yee}, {Ryu}, {Albrow},
  {Chung}, {Han}, {Kil Jung}, {Lee}, {Shin}, {Shvartzvald}, {Zang}, {Cha},
  {Kim}, {Kim}, {Lee}, {Lee}, {Park}, and {Pogge}}]{kim2021}
{Kim} HW, {Hwang} KH, {Gould} A et~al. (2021) {KMT-2019-BLG-2073: Fourth
  Free-floating Planet Candidate with {\ensuremath{\theta}}$_{E} < 10 $
  {\ensuremath{\mu}}as}. \aj 162(1):15

\bibitem[{{Kim} et~al.(2016){Kim}, {Lee}, {Park}, {Kim}, {Cha}, {Lee}, {Han},
  {Chun}, and {Yuk}}]{kim2016}
{Kim} SL, {Lee} CU, {Park} BG et~al. (2016) {KMTNET: A Network of 1.6 m
  Wide-Field Optical Telescopes Installed at Three Southern Observatories}.
  Journal of Korean Astronomical Society 49(1):37--44

\bibitem[{{Kirkpatrick} et~al.(2019){Kirkpatrick}, {Martin}, {Smart}, {Cayago},
  {Beichman}, {Marocco}, {Gelino}, {Faherty}, {Cushing}, {Schneider}, {Mace},
  {Tinney}, {Wright}, {Lowrance}, {Ingalls}, {Vrba}, {Munn}, {Dahm}, and
  {McLean}}]{kirkpatrick2019}
{Kirkpatrick} JD, {Martin} EC, {Smart} RL et~al. (2019) {Preliminary
  Trigonometric Parallaxes of 184 Late-T and Y Dwarfs and an Analysis of the
  Field Substellar Mass Function into the
  {\textquotedblleft}Planetary{\textquotedblright} Mass Regime}. \apjs
  240(2):19

\bibitem[{{Kirkpatrick} et~al.(2021){Kirkpatrick}, {Gelino}, {Faherty},
  {Meisner}, {Caselden}, {Schneider}, {Marocco}, {Cayago}, {Smart},
  {Eisenhardt}, {Kuchner}, {Wright}, {Cushing}, {Allers}, {Bardalez Gagliuffi},
  {Burgasser}, {Gagn{\'e}}, {Logsdon}, {Martin}, {Ingalls}, {Lowrance},
  {Abrahams}, {Aganze}, {Gerasimov}, {Gonzales}, {Hsu}, {Kamraj}, {Kiman},
  {Rees}, {Theissen}, {Ammar}, {Andersen}, {Beaulieu}, {Colin}, {Elachi},
  {Goodman}, {Gramaize}, {Hamlet}, {Hong}, {Jonkeren}, {Khalil}, {Martin},
  {Pendrill}, {Pumphrey}, {Rothermich}, {Sainio}, {Stenner}, {Tanner},
  {Th{\'e}venot}, {Voloshin}, {Walla}, {W{\k{e}}dracki}, and {Backyard Worlds:
  Planet 9 Collaboration}}]{kirkpatrick2021}
{Kirkpatrick} JD, {Gelino} CR, {Faherty} JK et~al. (2021) {The Field Substellar
  Mass Function Based on the Full-sky 20 pc Census of 525 L, T, and Y Dwarfs}.
  \apjs 253(1):7

\bibitem[{{Koshimoto} et~al.(2021){Koshimoto}, {Bennett}, {Suzuki}, and
  {Bond}}]{koshimoto2021}
{Koshimoto} N, {Bennett} DP, {Suzuki} D {Bond} IA (2021) {No Large Dependence
  of Planet Frequency on Galactocentric Distance}. \apjl 918(1):L8

\bibitem[{{Koshimoto} et~al.(2023){Koshimoto}, {Sumi}, {Bennett}, {Bozza},
  {Mr{\'o}z}, {Udalski}, {Rattenbury}, {Abe}, {Barry}, {Bhattacharya}, {Bond},
  {Fujii}, {Fukui}, {Hamada}, {Hirao}, {Silva}, {Itow}, {Kirikawa}, {Kondo},
  {Matsubara}, {Miyazaki}, {Muraki}, {Olmschenk}, {Ranc}, {Satoh}, {Suzuki},
  {Tomoyoshi}, {Tristram}, {Vandorou}, {Yama}, and {Yamashita}}]{koshimoto2023}
{Koshimoto} N, {Sumi} T, {Bennett} DP et~al. (2023) {Terrestrial- and
  Neptune-mass Free-Floating Planet Candidates from the MOA-II 9 yr Galactic
  Bulge Survey}. \aj 166(3):107

\bibitem[{{Kubas} et~al.(2008){Kubas}, {Cassan}, {Dominik}, {Bennett},
  {Wambsganss}, {Brillant}, {Beaulieu}, {Albrow}, {Batista}, {Bode}, {Bramich},
  {Burgdorf}, {Caldwell}, {Calitz}, {Cook}, {Coutures}, {Dieters}, {Dominis
  Prester}, {Donatowicz}, {Fouqu{\'e}}, {Greenhill}, {Hill}, {Hoffman},
  {Horne}, {J{\o}rgensen}, {Kains}, {Kane}, {Marquette}, {Martin}, {Meintjes},
  {Menzies}, {Pollard}, {Sahu}, {Snodgrass}, {Steele}, {Tsapras}, {Vinter},
  {Williams}, {Woller}, {Zub}, and {PLANET/RoboNet Collaboration}}]{kubas2008}
{Kubas} D, {Cassan} A, {Dominik} M et~al. (2008) {Limits on additional
  planetary companions to OGLE 2005-BLG-390L}. \aap 483(1):317--324

\bibitem[{{Lin} and {Ida}(1997)}]{lin1997}
{Lin} DNC {Ida} S (1997) {On the Origin of Massive Eccentric Planets}. \apj
  477(2):781--791

\bibitem[{{Lodieu} et~al.(2021){Lodieu}, {Hambly}, and {Cross}}]{lodieu2021}
{Lodieu} N, {Hambly} NC {Cross} NJG (2021) {Exploring the planetary-mass
  population in the Upper Scorpius association}. \mnras 503(2):2265--2279

\bibitem[{{Ma} et~al.(2016){Ma}, {Mao}, {Ida}, {Zhu}, and {Lin}}]{ma2016}
{Ma} S, {Mao} S, {Ida} S, {Zhu} W {Lin} DNC (2016) {Free-floating planets from
  core accretion theory: microlensing predictions}. \mnras 461(1):L107--L111

\bibitem[{{Madsen} and {Zhu}(2019)}]{madsen2019}
{Madsen} S {Zhu} W (2019) {A Pair of Planets Likely in Mean-motion Resonance
  From Gravitational Microlensing}. \apjl 878(2):L29

\bibitem[{{Malmberg} et~al.(2011){Malmberg}, {Davies}, and
  {Heggie}}]{malmberg2011}
{Malmberg} D, {Davies} MB {Heggie} DC (2011) {The effects of fly-bys on
  planetary systems}. \mnras 411(2):859--877

\bibitem[{{Mao} and {Paczy\'nski}(1991)}]{mao1991}
{Mao} S {Paczy\'nski} B (1991) {Gravitational Microlensing by Double Stars and
  Planetary Systems}. \apjl 374:L37

\bibitem[{{Matsumoto} et~al.(2020){Matsumoto}, {Gu}, {Kokubo}, {Oshino}, and
  {Omiya}}]{matsumoto2020}
{Matsumoto} Y, {Gu} PG, {Kokubo} E, {Oshino} S {Omiya} M (2020) {Ejection of
  close-in super-Earths around low-mass stars in the giant impact stage}. \aap
  642:A23

\bibitem[{{Matsumura} et~al.(2013){Matsumura}, {Ida}, and
  {Nagasawa}}]{matsumura2013}
{Matsumura} S, {Ida} S {Nagasawa} M (2013) {Effects of Dynamical Evolution of
  Giant Planets on Survival of Terrestrial Planets}. \apj 767(2):129

\bibitem[{{McDonald} et~al.(2021){McDonald}, {Kerins}, {Poleski}, {Penny},
  {Specht}, {Mao}, {Fouqu{\'e}}, {Zhu}, and {Zang}}]{mcdonald2021}
{McDonald} I, {Kerins} E, {Poleski} R et~al. (2021) {Kepler K2 Campaign 9 - I.
  Candidate short-duration events from the first space-based survey for
  planetary microlensing}. \mnras 505(4):5584--5602

\bibitem[{{Medford} et~al.(2023){Medford}, {Abrams}, {Lu}, {Nugent}, and
  {Lam}}]{medford2023}
{Medford} MS, {Abrams} NS, {Lu} JR, {Nugent} P {Lam} CY (2023) {60 Microlensing
  Events from the Three Years of Zwicky Transient Facility Phase One}. \apj
  947(1):24

\bibitem[{{Miret-Roig} et~al.(2022){Miret-Roig}, {Bouy}, {Raymond}, {Tamura},
  {Bertin}, {Barrado}, {Olivares}, {Galli}, {Cuillandre}, {Sarro}, {Berihuete},
  and {Hu{\'e}lamo}}]{miretroig2022}
{Miret-Roig} N, {Bouy} H, {Raymond} SN et~al. (2022) {A rich population of
  free-floating planets in the Upper Scorpius young stellar association}.
  Nature Astronomy 6:89--97

\bibitem[{{Mordasini} et~al.(2009){Mordasini}, {Alibert}, and
  {Benz}}]{mordasini2009}
{Mordasini} C, {Alibert} Y {Benz} W (2009) {Extrasolar planet population
  synthesis. I. Method, formation tracks, and mass-distance distribution}. \aap
  501(3):1139--1160

\bibitem[{{Mr{\'o}z} et~al.(2017){Mr{\'o}z}, {Udalski}, {Skowron}, {Poleski},
  {Koz{\l}owski}, {Szyma{\'n}ski}, {Soszy{\'n}ski}, {Wyrzykowski},
  {Pietrukowicz}, {Ulaczyk}, {Skowron}, and {Pawlak}}]{mroz2017}
{Mr{\'o}z} P, {Udalski} A, {Skowron} J et~al. (2017) {No large population of
  unbound or wide-orbit Jupiter-mass planets}. \nat 548(7666):183--186

\bibitem[{{Mr{\'o}z} et~al.(2018){Mr{\'o}z}, {Ryu}, {Skowron}, {Udalski},
  {Gould}, {Szyma{\'n}ski}, {Soszy{\'n}ski}, {Poleski}, {Pietrukowicz},
  {Koz{\l}owski}, {Pawlak}, {Ulaczyk}, {OGLE Collaboration}, {Albrow}, {Chung},
  {Jung}, {Han}, {Hwang}, {Shin}, {Yee}, {Zhu}, {Cha}, {Kim}, {Kim}, {Kim},
  {Lee}, {Lee}, {Lee}, {Park}, {Pogge}, and {KMTNet Collaboration}}]{mroz2018}
{Mr{\'o}z} P, {Ryu} YH, {Skowron} J et~al. (2018) {A Neptune-mass Free-floating
  Planet Candidate Discovered by Microlensing Surveys}. \aj 155(3):121

\bibitem[{{Mr{\'o}z} et~al.(2019){Mr{\'o}z}, {Udalski}, {Bennett}, {Ryu},
  {Sumi}, {Shvartzvald}, {Skowron}, {Poleski}, {Pietrukowicz}, {Koz{\l}owski},
  {Szyma{\'n}ski}, {Wyrzykowski}, {Soszy{\'n}ski}, {Ulaczyk}, {Rybicki},
  {Iwanek}, {Albrow}, {Chung}, {Gould}, {Han}, {Hwang}, {Jung}, {Shin}, {Yee},
  {Zang}, {Cha}, {Kim}, {Kim}, {Kim}, {Lee}, {Lee}, {Lee}, {Park}, {Pogge},
  {Abe}, {Barry}, {Bhattacharya}, {Bond}, {Donachie}, {Fukui}, {Hirao}, {Itow},
  {Kawasaki}, {Kondo}, {Koshimoto}, {Li}, {Matsubara}, {Muraki}, {Miyazaki},
  {Nagakane}, {Ranc}, {Rattenbury}, {Suematsu}, {Sullivan}, {Suzuki},
  {Tristram}, {Yonehara}, {Maoz}, {Kaspi}, and {Friedmann}}]{mroz2019}
{Mr{\'o}z} P, {Udalski} A, {Bennett} DP et~al. (2019) {Two new free-floating or
  wide-orbit planets from microlensing}. \aap 622:A201

\bibitem[{{Mr{\'o}z} et~al.(2020{\natexlab{a}}){Mr{\'o}z}, {Poleski}, {Gould},
  {Udalski}, {Sumi}, {Szyma{\'n}ski}, {Soszy{\'n}ski}, {Pietrukowicz},
  {Koz{\l}owski}, {Skowron}, {Ulaczyk}, {OGLE Collaboration}, {Albrow},
  {Chung}, {Han}, {Hwang}, {Jung}, {Kim}, {Ryu}, {Shin}, {Shvartzvald}, {Yee},
  {Zang}, {Cha}, {Kim}, {Kim}, {Lee}, {Lee}, {Lee}, {Park}, {Pogge}, and {KMT
  Collaboration}}]{mroz2020a}
{Mr{\'o}z} P, {Poleski} R, {Gould} A et~al. (2020{\natexlab{a}}) {A
  Terrestrial-mass Rogue Planet Candidate Detected in the Shortest-timescale
  Microlensing Event}. \apjl 903(1):L11

\bibitem[{{Mr{\'o}z} et~al.(2020{\natexlab{b}}){Mr{\'o}z}, {Poleski}, {Han},
  {Udalski}, {Gould}, {Szyma{\'n}ski}, {Soszy{\'n}ski}, {Pietrukowicz},
  {Koz{\l}owski}, {Skowron}, {Ulaczyk}, {Gromadzki}, {Rybicki}, {Iwanek},
  {Wrona}, {OGLE Collaboration}, {Albrow}, {Chung}, {Hwang}, {Ryu}, {Jung},
  {Shin}, {Shvartzvald}, {Yee}, {Zang}, {Cha}, {Kim}, {Kim}, {Kim}, {Lee},
  {Lee}, {Lee}, {Park}, {Pogge}, and {KMT Collaboration}}]{mroz2020b}
{Mr{\'o}z} P, {Poleski} R, {Han} C et~al. (2020{\natexlab{b}}) {A Free-floating
  or Wide-orbit Planet in the Microlensing Event OGLE-2019-BLG-0551}. \aj
  159(6):262

\bibitem[{{Mr\'oz} et~al.(2023){Mr\'oz}, {Ban}, {Marty}, and
  {Poleski}}]{mroz2023}
{Mr\'oz} P, {Ban} M, {Marty} P {Poleski} R (2023) {Free-floating or wide-orbit?
  Keck adaptive-optics observations reveal no host stars near free-floating
  planet candidates}. arXiv e-prints arXiv:2303.04610

\bibitem[{{Nemiroff} and {Wickramasinghe}(1994)}]{nemiroff1994}
{Nemiroff} RJ {Wickramasinghe} WADT (1994) {Finite Source Sizes and the
  Information Content of Macho-Type Lens Search Light Curves}. \apjl 424:L21

\bibitem[{{Paczy\'nski}(1986)}]{paczynski1986}
{Paczy\'nski} B (1986) {Gravitational Microlensing by the Galactic Halo}. \apj
  304:1

\bibitem[{{Padoan} and {Nordlund}(2002)}]{padoan2002}
{Padoan} P {Nordlund} {\r{A}} (2002) {The Stellar Initial Mass Function from
  Turbulent Fragmentation}. \apj 576(2):870--879

\bibitem[{{Parker} and {Quanz}(2012)}]{parker2012}
{Parker} RJ {Quanz} SP (2012) {The effects of dynamical interactions on planets
  in young substructured star clusters}. \mnras 419(3):2448--2458

\bibitem[{{Pe{\~n}a Ram{\'\i}rez} et~al.(2012){Pe{\~n}a Ram{\'\i}rez},
  {B{\'e}jar}, {Zapatero Osorio}, {Petr-Gotzens}, and {Mart{\'\i}n}}]{pena2012}
{Pe{\~n}a Ram{\'\i}rez} K, {B{\'e}jar} VJS, {Zapatero Osorio} MR,
  {Petr-Gotzens} MG {Mart{\'\i}n} EL (2012) {New Isolated Planetary-mass
  Objects and the Stellar and Substellar Mass Function of the
  {\ensuremath{\sigma}} Orionis Cluster}. \apj 754(1):30

\bibitem[{{Penny} et~al.(2016){Penny}, {Henderson}, and {Clanton}}]{penny2016}
{Penny} MT, {Henderson} CB {Clanton} C (2016) {Is the Galactic Bulge Devoid of
  Planets?} \apj 830(2):150

\bibitem[{{Penny} et~al.(2019){Penny}, {Gaudi}, {Kerins}, {Rattenbury}, {Mao},
  {Robin}, and {Calchi Novati}}]{penny2019}
{Penny} MT, {Gaudi} BS, {Kerins} E et~al. (2019) {Predictions of the WFIRST
  Microlensing Survey. I. Bound Planet Detection Rates}. \apjs 241(1):3

\bibitem[{{Pfyffer} et~al.(2015){Pfyffer}, {Alibert}, {Benz}, and
  {Swoboda}}]{pfyffer2015}
{Pfyffer} S, {Alibert} Y, {Benz} W {Swoboda} D (2015) {Theoretical models of
  planetary system formation. II. Post-formation evolution}. \aap 579:A37

\bibitem[{{Poleski} et~al.(2014){Poleski}, {Skowron}, {Udalski}, {Han},
  {Koz{\l}owski}, {Wyrzykowski}, {Dong}, {Szyma{\'n}ski}, {Kubiak},
  {Pietrzy{\'n}ski}, {Soszy{\'n}ski}, {Ulaczyk}, {Pietrukowicz}, and
  {Gould}}]{poleski2014c}
{Poleski} R, {Skowron} J, {Udalski} A et~al. (2014) {Triple Microlens
  OGLE-2008-BLG-092L: Binary Stellar System with a Circumprimary Uranus-type
  Planet}. \apj 795(1):42

\bibitem[{{Poleski} et~al.(2021){Poleski}, {Skowron}, {Mr{\'o}z}, {Udalski},
  {Szyma{\'n}ski}, {Pietrukowicz}, {Ulaczyk}, {Rybicki}, {Iwanek}, {Wrona}, and
  {Gromadzki}}]{poleski21}
{Poleski} R, {Skowron} J, {Mr{\'o}z} P et~al. (2021) {Wide-Orbit Exoplanets are
  Common. Analysis of Nearly 20 Years of OGLE Microlensing Survey Data}. \actaa
  71(1):1--23

\bibitem[{{Pollack} et~al.(1996){Pollack}, {Hubickyj}, {Bodenheimer},
  {Lissauer}, {Podolak}, and {Greenzweig}}]{pollack1996}
{Pollack} JB, {Hubickyj} O, {Bodenheimer} P et~al. (1996) {Formation of the
  Giant Planets by Concurrent Accretion of Solids and Gas}. \icarus
  124(1):62--85

\bibitem[{{Rasio} and {Ford}(1996)}]{rasio1996}
{Rasio} FA {Ford} EB (1996) {Dynamical instabilities and the formation of
  extrasolar planetary systems}. Science 274:954--956

\bibitem[{{Raymond} and {Morbidelli}(2022)}]{raymond2022}
{Raymond} SN {Morbidelli} A (2022) {Planet Formation: Key Mechanisms and Global
  Models}. In: {Biazzo} K, {Bozza} V, {Mancini} L {Sozzetti} A (eds)
  Demographics of Exoplanetary Systems, Lecture Notes of the 3rd Advanced
  School on Exoplanetary Science, Astrophysics and Space Science Library, vol
  466, pp 3--82, \doi{10.1007/978-3-030-88124-5_1}

\bibitem[{{Reipurth} and {Clarke}(2001)}]{reipurth2001}
{Reipurth} B {Clarke} C (2001) {The Formation of Brown Dwarfs as Ejected
  Stellar Embryos}. \aj 122(1):432--439

\bibitem[{{Rhie} et~al.(2000){Rhie}, {Bennett}, {Becker}, {Peterson},
  {Fragile}, {Johnson}, {Quinn}, {Crouch}, {Gray}, {King}, {Messenger},
  {Thomson}, {Bond}, {Abe}, {Carter}, {Dodd}, {Hearnshaw}, {Honda}, {Jugaku},
  {Kabe}, {Kilmartin}, {Koribalski}, {Masuda}, {Matsubara}, {Muraki},
  {Nakamura}, {Nankivell}, {Noda}, {Rattenbury}, {Reid}, {Rumsey}, {Saito},
  {Sato}, {Sato}, {Sekiguchi}, {Sullivan}, {Sumi}, {Watase}, {Yanagisawa},
  {Yock}, and {Yoshizawa}}]{rhie00}
{Rhie} SH, {Bennett} DP, {Becker} AC et~al. (2000) {On Planetary Companions to
  the MACHO 98-BLG-35 Microlens Star}. \apj 533(1):378--391

\bibitem[{{Rickman} et~al.(2023){Rickman}, {Wajer}, {Przy{\l}uski},
  {Wi{\'s}niowski}, {Nesvorn{\'y}}, and {Morbidelli}}]{rickman2023}
{Rickman} H, {Wajer} P, {Przy{\l}uski} R et~al. (2023) {Breakdown of planetary
  systems in embedded clusters}. \mnras 520(1):637--648

\bibitem[{{Rodriguez} et~al.(2022){Rodriguez}, {Mr{\'o}z}, {Kulkarni},
  {Andreoni}, {Bellm}, {Dekany}, {Drake}, {Duev}, {Graham}, {Masci}, {Prince},
  {Riddle}, and {Shupe}}]{rodriguez2022}
{Rodriguez} AC, {Mr{\'o}z} P, {Kulkarni} SR et~al. (2022) {Microlensing Events
  in the Galactic Plane Using the Zwicky Transient Facility}. \apj 927(2):150

\bibitem[{{Ryu} et~al.(2021){Ryu}, {Mr{\'o}z}, {Gould}, {Hwang}, {Kim}, {Yee},
  {Albrow}, {Chung}, {Jung}, {Shin}, {Shvartzvald}, {Zang}, {Cha}, {Kim},
  {Kim}, {Lee}, {Lee}, {Lee}, {Park}, {Han}, {Pogge}, {KMTNet Collaboration},
  {Udalski}, {Poleski}, {Skowron}, {Szyma{\'n}ski}, {Soszy{\'n}ski},
  {Pietrukowicz}, {Koz{\l}owski}, {Ulaczyk}, {Rybicki}, {Iwanek}, and {OGLE
  Collaboration}}]{ryu2021}
{Ryu} YH, {Mr{\'o}z} P, {Gould} A et~al. (2021) {KMT-2017-BLG-2820 and the
  Nature of the Free-floating Planet Population}. \aj 161(3):126

\bibitem[{{Ryu} et~al.(2023){Ryu}, {Udalski}, {Yee}, {Zang}, {Shvartzvald},
  {Han}, {Gould}, {Albrow}, {Chung}, {Hwang}, {Jung}, {Shin}, {Yang}, {Cha},
  {Kim}, {Kim}, {Lee}, {Lee}, {Lee}, {Park}, {Pogge}, {Wang}, {Mr{\'o}z},
  {Szyma{\'n}ski}, {Skowron}, {Poleski}, {Soszy{\'n}ski}, {Pietrukowicz},
  {Koz{\l}owski}, {Ulaczyk}, {Rybicki}, {Iwanek}, {Wrona}, {Beichman},
  {Bryden}, {Carey}, {Henderson}, {Calchi Novati}, {Zhu}, {Jacklin}, and
  {Penny}}]{ryu2023a}
{Ryu} YH, {Udalski} A, {Yee} JC et~al. (2023) {Systematic KMTNet Planetary
  Anomaly Search. X. Complete Sample of 2017 Prime-Field Planets}. arXiv
  e-prints arXiv:2307.13359

\bibitem[{{Sahu} et~al.(2001){Sahu}, {Casertano}, {Livio}, {Gilliland},
  {Panagia}, {Albrow}, and {Potter}}]{sahu2001}
{Sahu} KC, {Casertano} S, {Livio} M et~al. (2001) {Gravitational microlensing
  by low-mass objects in the globular cluster M22}. \nat 411(6841):1022--1024

\bibitem[{{Sahu} et~al.(2002){Sahu}, {Anderson}, and {King}}]{sahu2002}
{Sahu} KC, {Anderson} J {King} IR (2002) {A Reexamination of the ``Planetary''
  Lensing Events in M22}. \apjl 565(1):L21--L24

\bibitem[{{Schmidt}(1968)}]{schmidt1968}
{Schmidt} M (1968) {Space Distribution and Luminosity Functions of
  Quasi-Stellar Radio Sources}. \apj 151:393

\bibitem[{{Scholz} et~al.(2012){Scholz}, {Jayawardhana}, {Muzic}, {Geers},
  {Tamura}, and {Tanaka}}]{scholz2012}
{Scholz} A, {Jayawardhana} R, {Muzic} K et~al. (2012) {Substellar Objects in
  Nearby Young Clusters (SONYC). VI. The Planetary-mass Domain of NGC 1333}.
  \apj 756(1):24

\bibitem[{{Shin} et~al.(2023){Shin}, {Yee}, {Zang}, {Yang}, {Hwang}, {Han},
  {Gould}, {Udalski}, {Bond}, {Albrow}, {Chung}, {Jung}, {Ryu}, {Shvartzvald},
  {Cha}, {Kim}, {Kim}, {Lee}, {Lee}, {Lee}, {Park}, {Pogge}, {Mr{\'o}z},
  {Szyma{\'n}ski}, {Skowron}, {Poleski}, {Soszy{\'n}ski}, {Pietrukowicz},
  {Koz{\l}owski}, {Rybicki}, {Iwanek}, {Ulaczyk}, {Wrona}, {Gromadzki}, {Abe},
  {Barry}, {Bennett}, {Bhattacharya}, {Fujii}, {Fukui}, {Hamada}, {Hirao},
  {Silva}, {Itow}, {Kirikawa}, {Kondo}, {Koshimoto}, {Matsubara}, {Miyazaki},
  {Muraki}, {Olmschenk}, {Ranc}, {Rattenbury}, {Satoh}, {Sumi}, {Suzuki},
  {Tomoyoshi}, {Tristram}, {Vandorou}, {Yama}, and {Yamashita}}]{shin2023a}
{Shin} IG, {Yee} JC, {Zang} W et~al. (2023) {Systematic KMTNet Planetary
  Anomaly Search. IX. Complete Sample of 2016 Prime-field Planets}. \aj
  166(3):104

\bibitem[{{Shvartzvald} et~al.(2016){Shvartzvald}, {Maoz}, {Udalski}, {Sumi},
  {Friedmann}, {Kaspi}, {Poleski}, {Szyma{\'n}ski}, {Skowron}, {Koz{\l}owski},
  {Wyrzykowski}, {Mr{\'o}z}, {Pietrukowicz}, {Pietrzy{\'n}ski},
  {Soszy{\'n}ski}, {Ulaczyk}, {Abe}, {Barry}, {Bennett}, {Bhattacharya},
  {Bond}, {Freeman}, {Inayama}, {Itow}, {Koshimoto}, {Ling}, {Masuda}, {Fukui},
  {Matsubara}, {Muraki}, {Ohnishi}, {Rattenbury}, {Saito}, {Sullivan},
  {Suzuki}, {Tristram}, {Wakiyama}, and {Yonehara}}]{shvartzvald16}
{Shvartzvald} Y, {Maoz} D, {Udalski} A et~al. (2016) {The frequency of
  snowline-region planets from four years of OGLE-MOA-Wise second-generation
  microlensing}. \mnras 457(4):4089--4113

\bibitem[{{Skowron} et~al.(2011){Skowron}, {Udalski}, {Gould}, {Dong},
  {Monard}, {Han}, {Nelson}, {McCormick}, {Moorhouse}, {Thornley}, {Maury},
  {Bramich}, {Greenhill}, {Koz{\l}owski}, {Bond}, {Poleski}, {Wyrzykowski},
  {Ulaczyk}, {Kubiak}, {Szyma{\'n}ski}, {Pietrzy{\'n}ski}, {Soszy{\'n}ski},
  {OGLE Collaboration}, {Gaudi}, {Yee}, {Hung}, {Pogge}, {DePoy}, {Lee},
  {Park}, {Allen}, {Mallia}, {Drummond}, {Bolt}, {{\ensuremath{\mu}}FUN
  Collaboration}, {Allan}, {Browne}, {Clay}, {Dominik}, {Fraser}, {Horne},
  {Kains}, {Mottram}, {Snodgrass}, {Steele}, {Street}, {Tsapras}, {RoboNet
  Collaboration}, {Abe}, {Bennett}, {Botzler}, {Douchin}, {Freeman}, {Fukui},
  {Furusawa}, {Hayashi}, {Hearnshaw}, {Hosaka}, {Itow}, {Kamiya}, {Kilmartin},
  {Korpela}, {Lin}, {Ling}, {Makita}, {Masuda}, {Matsubara}, {Muraki},
  {Nagayama}, {Miyake}, {Nishimoto}, {Ohnishi}, {Perrott}, {Rattenbury},
  {Saito}, {Skuljan}, {Sullivan}, {Sumi}, {Suzuki}, {Sweatman}, {Tristram},
  {Wada}, {Yock}, {MOA Collaboration}, {Beaulieu}, {Fouqu{\'e}}, {Albrow},
  {Batista}, {Brillant}, {Caldwell}, {Cassan}, {Cole}, {Cook}, {Coutures},
  {Dieters}, {Dominis Prester}, {Donatowicz}, {Kane}, {Kubas}, {Marquette},
  {Martin}, {Menzies}, {Sahu}, {Wambsganss}, {Williams}, {Zub}, and {PLANET
  Collaboration}}]{skowron2011}
{Skowron} J, {Udalski} A, {Gould} A et~al. (2011) {Binary Microlensing Event
  OGLE-2009-BLG-020 Gives Verifiable Mass, Distance, and Orbit Predictions}.
  \apj 738(1):87

\bibitem[{{Snodgrass} et~al.(2004){Snodgrass}, {Horne}, and
  {Tsapras}}]{snodgrass2004}
{Snodgrass} C, {Horne} K {Tsapras} Y (2004) {The abundance of Galactic planets
  from OGLE-III 2002 microlensing data}. \mnras 351(3):967--975

\bibitem[{{Spurzem} et~al.(2009){Spurzem}, {Giersz}, {Heggie}, and
  {Lin}}]{spurzem2009}
{Spurzem} R, {Giersz} M, {Heggie} DC {Lin} DNC (2009) {Dynamics of Planetary
  Systems in Star Clusters}. \apj 697(1):458--482

\bibitem[{{Street} et~al.(2018){Street}, {Lund}, {Donachie}, {Khakpash},
  {Golovich}, {Penny}, {Bennett}, {Dawson}, {Pepper}, {Rabus}, {Szkody},
  {Clarkson}, {Di Stefano}, {Rattenbury}, {Hundertmark}, {Tsapras}, {Ridgway},
  {Stassun}, {Bozza}, {Bhattacharya}, {Calchi Novati}, and
  {Shvartzvald}}]{street2018}
{Street} RA, {Lund} MB, {Donachie} M et~al. (2018) {Unique Science from a
  Coordinated LSST-WFIRST Survey of the Galactic Bulge}. arXiv e-prints
  arXiv:1812.04445

\bibitem[{{Sumi} et~al.(2003){Sumi}, {Abe}, {Bond}, {Dodd}, {Hearnshaw},
  {Honda}, {Honma}, {Kan-ya}, {Kilmartin}, {Masuda}, {Matsubara}, {Muraki},
  {Nakamura}, {Nishi}, {Noda}, {Ohnishi}, {Petterson}, {Rattenbury}, {Reid},
  {Saito}, {Saito}, {Sato}, {Sekiguchi}, {Skuljan}, {Sullivan}, {Takeuti},
  {Tristram}, {Wilkinson}, {Yanagisawa}, and {Yock}}]{sumi2003}
{Sumi} T, {Abe} F, {Bond} IA et~al. (2003) {Microlensing Optical Depth toward
  the Galactic Bulge from Microlensing Observations in Astrophysics Group
  Observations during 2000 with Difference Image Analysis}. \apj
  591(1):204--227

\bibitem[{{Sumi} et~al.(2010){Sumi}, {Bennett}, {Bond}, {Udalski}, {Batista},
  {Dominik}, {Fouqu{\'e}}, {Kubas}, {Gould}, {Macintosh}, {Cook}, {Dong},
  {Skuljan}, {Cassan}, {Abe}, {Botzler}, {Fukui}, {Furusawa}, {Hearnshaw},
  {Itow}, {Kamiya}, {Kilmartin}, {Korpela}, {Lin}, {Ling}, {Masuda},
  {Matsubara}, {Miyake}, {Muraki}, {Nagaya}, {Nagayama}, {Ohnishi}, {Okumura},
  {Perrott}, {Rattenbury}, {Saito}, {Sako}, {Sullivan}, {Sweatman}, {Tristram},
  {Yock}, {MOA Collaboration}, {Beaulieu}, {Cole}, {Coutures}, {Duran},
  {Greenhill}, {Jablonski}, {Marboeuf}, {Martioli}, {Pedretti}, {Pejcha},
  {Rojo}, {Albrow}, {Brillant}, {Bode}, {Bramich}, {Burgdorf}, {Caldwell},
  {Calitz}, {Corrales}, {Dieters}, {Dominis Prester}, {Donatowicz}, {Hill},
  {Hoffman}, {Horne}, {J{\o}rgensen}, {Kains}, {Kane}, {Marquette}, {Martin},
  {Meintjes}, {Menzies}, {Pollard}, {Sahu}, {Snodgrass}, {Steele}, {Street},
  {Tsapras}, {Wambsganss}, {Williams}, {Zub}, {PLANET Collaboration},
  {Szyma{\'n}ski}, {Kubiak}, {Pietrzy{\'n}ski}, {Soszy{\'n}ski}, {Szewczyk},
  {Wyrzykowski}, {Ulaczyk}, {OGLE Collaboration}, {Allen}, {Christie}, {DePoy},
  {Gaudi}, {Han}, {Janczak}, {Lee}, {McCormick}, {Mallia}, {Monard}, {Natusch},
  {Park}, {Pogge}, {Santallo}, and {{\ensuremath{\mu}}FUN
  Collaboration}}]{sumi10}
{Sumi} T, {Bennett} DP, {Bond} IA et~al. (2010) {A Cold Neptune-Mass Planet
  OGLE-2007-BLG-368Lb: Cold Neptunes Are Common}. \apj 710(2):1641--1653

\bibitem[{{Sumi} et~al.(2011){Sumi}, {Kamiya}, {Bennett}, {Bond}, {Abe},
  {Botzler}, {Fukui}, {Furusawa}, {Hearnshaw}, {Itow}, {Kilmartin}, {Korpela},
  {Lin}, {Ling}, {Masuda}, {Matsubara}, {Miyake}, {Motomura}, {Muraki},
  {Nagaya}, {Nakamura}, {Ohnishi}, {Okumura}, {Perrott}, {Rattenbury}, {Saito},
  {Sako}, {Sullivan}, {Sweatman}, {Tristram}, {Udalski}, {Szyma{\'n}ski},
  {Kubiak}, {Pietrzy{\'n}ski}, {Poleski}, {Soszy{\'n}ski}, {Wyrzykowski},
  {Ulaczyk}, and {Microlensing Observations in Astrophysics (MOA)
  Collaboration}}]{sumi2011}
{Sumi} T, {Kamiya} K, {Bennett} DP et~al. (2011) {Unbound or distant planetary
  mass population detected by gravitational microlensing}. \nat
  473(7347):349--352

\bibitem[{{Sumi} et~al.(2023){Sumi}, {Koshimoto}, {Bennett}, {Rattenbury},
  {Abe}, {Barry}, {Bhattacharya}, {Bond}, {Fujii}, {Fukui}, {Hamada}, {Hirao},
  {Silva}, {Itow}, {Kirikawa}, {Kondo}, {Matsubara}, {Miyazaki}, {Muraki},
  {Olmschenk}, {Ranc}, {Satoh}, {Suzuki}, {Tomoyoshi}, {Tristram}, {Vandorou},
  {Yama}, and {Yamashita}}]{sumi2023}
{Sumi} T, {Koshimoto} N, {Bennett} DP et~al. (2023) {Free-floating Planet Mass
  Function from MOA-II 9 yr Survey toward the Galactic Bulge}. \aj 166(3):108

\bibitem[{{Sutherland} and {Fabrycky}(2016)}]{sutherland2016}
{Sutherland} AP {Fabrycky} DC (2016) {On the Fate of Unstable Circumbinary
  Planets: Tatooine{\textquoteright}s Close Encounters with a Death Star}. \apj
  818(1):6

\bibitem[{{Suzuki} et~al.(2016){Suzuki}, {Bennett}, {Sumi}, {Bond}, {Rogers},
  {Abe}, {Asakura}, {Bhattacharya}, {Donachie}, {Freeman}, {Fukui}, {Hirao},
  {Itow}, {Koshimoto}, {Li}, {Ling}, {Masuda}, {Matsubara}, {Muraki},
  {Nagakane}, {Onishi}, {Oyokawa}, {Rattenbury}, {Saito}, {Sharan}, {Shibai},
  {Sullivan}, {Tristram}, {Yonehara}, and {MOA Collaboration}}]{suzuki2016}
{Suzuki} D, {Bennett} DP, {Sumi} T et~al. (2016) {The Exoplanet Mass-ratio
  Function from the MOA-II Survey: Discovery of a Break and Likely Peak at a
  Neptune Mass}. \apj 833(2):145

\bibitem[{{Suzuki} et~al.(2018){Suzuki}, {Bennett}, {Ida}, {Mordasini},
  {Bhattacharya}, {Bond}, {Donachie}, {Fukui}, {Hirao}, {Koshimoto},
  {Miyazaki}, {Nagakane}, {Ranc}, {Rattenbury}, {Sumi}, {Alibert}, and
  {Lin}}]{suzuki2018b}
{Suzuki} D, {Bennett} DP, {Ida} S et~al. (2018) {Microlensing Results Challenge
  the Core Accretion Runaway Growth Scenario for Gas Giants}. \apjl 869(2):L34

\bibitem[{{Tsapras} et~al.(2016){Tsapras}, {Hundertmark}, {Wyrzykowski},
  {Horne}, {Udalski}, {Snodgrass}, {Street}, {Bramich}, {Dominik}, {Bozza},
  {Figuera Jaimes}, {Kains}, {Skowron}, {Szyma{\'n}ski}, {Pietrzy{\'n}ski},
  {Soszy{\'n}ski}, {Ulaczyk}, {Koz{\l}owski}, {Pietrukowicz}, and
  {Poleski}}]{tsapras16}
{Tsapras} Y, {Hundertmark} M, {Wyrzykowski} {\L} et~al. (2016) {The OGLE-III
  planet detection efficiency from six years of microlensing observations
  (2003-2008)}. \mnras 457(2):1320--1331

\bibitem[{{Udalski} et~al.(2015){Udalski}, {Szyma{\'n}ski}, and
  {Szyma{\'n}ski}}]{udalski2015b}
{Udalski} A, {Szyma{\'n}ski} MK {Szyma{\'n}ski} G (2015) {OGLE-IV: Fourth Phase
  of the Optical Gravitational Lensing Experiment}. \actaa 65(1):1--38

\bibitem[{{Udalski} et~al.(2018){Udalski}, {Ryu}, {Sajadian}, {Gould},
  {Mr{\'o}z}, {Poleski}, {Szyma{\'n}ski}, {Skowron}, {Soszy{\'n}ski},
  {Koz{\l}owski}, {Pietrukowicz}, {Ulaczyk}, {Pawlak}, {Rybicki}, {Iwanek},
  {Albrow}, {Chung}, {Han}, {Hwang}, {Jung}, {Shin}, {Shvartzvald}, {Yee},
  {Zang}, {Zhu}, {Cha}, {Kim}, {Kim}, {Kim}, {Lee}, {Lee}, {Lee}, {Park},
  {Pogge}, {Bozza}, {Dominik}, {Helling}, {Hundertmark}, {J{\o}rgensen},
  {Longa-Pe{\~n}a}, {Lowry}, {Burgdorf}, {Campbell-White}, {Ciceri}, {Evans},
  {Figuera Jaimes}, {Fujii}, {Haikala}, {Henning}, {Hinse}, {Mancini},
  {Peixinho}, {Rahvar}, {Rabus}, {Skottfelt}, {Snodgrass}, {Southworth}, and
  {von Essen}}]{udalski18}
{Udalski} A, {Ryu} YH, {Sajadian} S et~al. (2018) {OGLE-2017-BLG-1434Lb: Eighth
  $q<1\times 10^{-4}$ Mass-Ratio Microlens Planet Confirms Turnover in Planet
  Mass-Ratio Function}. \actaa 68(1):1--42

\bibitem[{{van Elteren} et~al.(2019){van Elteren}, {Portegies Zwart},
  {Pelupessy}, {Cai}, and {McMillan}}]{vanelteren2019}
{van Elteren} A, {Portegies Zwart} S, {Pelupessy} I, {Cai} MX {McMillan} SLW
  (2019) {Survivability of planetary systems in young and dense star clusters}.
  \aap 624:A120

\bibitem[{{Veras} and {Armitage}(2005)}]{veras2005}
{Veras} D {Armitage} PJ (2005) {The Influence of Massive Planet Scattering on
  Nascent Terrestrial Planets}. \apjl 620(2):L111--L114

\bibitem[{{Veras} and {Raymond}(2012)}]{veras2012}
{Veras} D {Raymond} SN (2012) {Planet-planet scattering alone cannot explain
  the free-floating planet population}. \mnras 421(1):L117--L121

\bibitem[{{Veras} et~al.(2011){Veras}, {Wyatt}, {Mustill}, {Bonsor}, and
  {Eldridge}}]{veras2011}
{Veras} D, {Wyatt} MC, {Mustill} AJ, {Bonsor} A {Eldridge} JJ (2011) {The great
  escape: how exoplanets and smaller bodies desert dying stars}. \mnras
  417(3):2104--2123

\bibitem[{{Wang} et~al.(2022){Wang}, {Zang}, {Zhu}, {Hwang}, {Udalski},
  {Gould}, {Han}, {Albrow}, {Chung}, {Jung}, {Kim}, {Ryu}, {Shin},
  {Shvartzvald}, {Yee}, {Cha}, {Kim}, {Kim}, {Kim}, {Lee}, {Lee}, {Lee},
  {Park}, {Pogge}, {Poleski}, {Mr{\'o}z}, {Skowron}, {Szyma{\'n}ski},
  {Soszy{\'n}ski}, {Pietrukowicz}, {Koz{\l}owski}, {Ulaczyk}, {Rybicki},
  {Iwanek}, {Wrona}, {Gromadzki}, {Yang}, {Mao}, and {Zhang}}]{wang2022a}
{Wang} H, {Zang} W, {Zhu} W et~al. (2022) {Systematic Korea Microlensing
  Telescope Network planetary anomaly search - III. One wide-orbit planet and
  two stellar binaries}. \mnras 510(2):1778--1790

\bibitem[{{Weidenschilling} and {Marzari}(1996)}]{weidenschilling1996}
{Weidenschilling} SJ {Marzari} F (1996) {Gravitational scattering as a possible
  origin for giant planets at small stellar distances}. \nat 384(6610):619--621

\bibitem[{{Whitworth} and {Zinnecker}(2004)}]{whitworth2004}
{Whitworth} AP {Zinnecker} H (2004) {The formation of free-floating brown
  dwarves and planetary-mass objects by photo-erosion of prestellar cores}.
  \aap 427:299--306

\bibitem[{{Witt} and {Mao}(1994)}]{witt1994}
{Witt} HJ {Mao} S (1994) {Can Lensed Stars Be Regarded as Pointlike for
  Microlensing by MACHOs?} \apj 430:505

\bibitem[{{Wu} et~al.(2023){Wu}, {Dong}, {Yi}, {Liu}, {El-Badry}, {Gould},
  {Christie}, {de Almeida}, {Monard}, {McCormick}, {Chen}, {Huang}, {Liu},
  {Merand}, {Mroz}, {Shangguan}, {Udalski}, {Woillez}, and {Zhang}}]{wu2023}
{Wu} Z, {Dong} S, {Yi} T et~al. (2023) {Gaia22dkvLb: A Microlensing Planet
  Potentially Accessible to Radial-Velocity Characterization}. arXiv e-prints
  arXiv:2309.03944

\bibitem[{{Wyrzykowski} et~al.(2020){Wyrzykowski}, {Mr{\'o}z}, {Rybicki},
  {Gromadzki}, {Ko{\l}aczkowski}, {Zieli{\'n}ski}, {Zieli{\'n}ski},
  {Britavskiy}, {Gomboc}, {Sokolovsky}, {Hodgkin}, {Abe}, {Aldi}, {AlMannaei},
  {Altavilla}, {Al Qasim}, {Anupama}, {Awiphan}, {Bachelet}, {Bak{\i}{\c{s}}},
  {Baker}, {Bartlett}, {Bendjoya}, {Benson}, {Bikmaev}, {Birenbaum},
  {Blagorodnova}, {Blanco-Cuaresma}, {Boeva}, {Bonanos}, {Bozza}, {Bramich},
  {Bruni}, {Burenin}, {Burgaz}, {Butterley}, {Caines}, {Caton}, {Calchi
  Novati}, {Carrasco}, {Cassan}, {{\v{C}}epas}, {Cropper},
  {Chru{\'s}li{\'n}ska}, {Clementini}, {Clerici}, {Conti}, {Conti}, {Cross},
  {Cusano}, {Damljanovic}, {Dapergolas}, {D'Ago}, {de Bruijne}, {Dennefeld},
  {Dhillon}, {Dominik}, {Dziedzic}, {Erece}, {Eselevich}, {Esenoglu}, {Eyer},
  {Figuera Jaimes}, {Fossey}, {Galeev}, {Grebenev}, {Gupta}, {Gutaev},
  {Hallakoun}, {Hamanowicz}, {Han}, {Handzlik}, {Haislip}, {Hanlon}, {Hardy},
  {Harrison}, {van Heerden}, {Hoette}, {Horne}, {Hudec}, {Hundertmark},
  {Ihanec}, {Irtuganov}, {Itoh}, {Iwanek}, {Jovanovic}, {Janulis},
  {Jel{\'\i}nek}, {Jensen}, {Kaczmarek}, {Katz}, {Khamitov}, {Kilic},
  {Klencki}, {Kolb}, {Kopacki}, {Kouprianov}, {Kruszy{\'n}ska}, {Kurowski},
  {Latev}, {Lee}, {Leonini}, {Leto}, {Lewis}, {Li}, {Liakos}, {Littlefair},
  {Lu}, {Manser}, {Mao}, {Maoz}, {Martin-Carrillo}, {Marais},
  {Maskoli{\={u}}nas}, {Maund}, {Meintjes}, {Melnikov}, {Ment},
  {Miko{\l}ajczyk}, {Morrell}, {Mowlavi}, {Mo{\'z}dzierski}, {Murphy},
  {Nazarov}, {Netzel}, {Nesci}, {Ngeow}, {Norton}, {Ofek},
  {Pak{\v{s}}tien{\.{e}}}, {Palaversa}, {Pandey}, {Paraskeva}, {Pawlak},
  {Penny}, {Penprase}, {Piascik}, {Prieto}, {Qvam}, {Ranc},
  {Rebassa-Mansergas}, {Reichart}, {Reig}, {Rhodes}, {Rivet}, {Rixon},
  {Roberts}, {Rosi}, {Russell}, {Zanmar Sanchez}, {Scarpetta}, {Seabroke},
  {Shappee}, {Schmidt}, {Shvartzvald}, {Sitek}, {Skowron}, {{\'S}niegowska},
  {Snodgrass}, {Soares}, {van Soelen}, {Spetsieri},
  {Stankevi{\v{c}}i{\={u}}t{\.{e}}}, {Steele}, {Street}, {Strobl}, {Strubble},
  {Szegedi}, {Tinjaca Ramirez}, {Tomasella}, {Tsapras}, {Vernet}, {Villanueva},
  {Vince}, {Wambsganss}, {van der Westhuizen}, {Wiersema}, {Wium}, {Wilson},
  {Yoldas}, {Zhuchkov}, {Zhukov}, {Zdanavi{\v{c}}ius}, {Zo{\l}a}, and
  {Zubareva}}]{wyrzykowski2020}
{Wyrzykowski} {\L}, {Mr{\'o}z} P, {Rybicki} KA et~al. (2020) {Full orbital
  solution for the binary system in the northern Galactic disc microlensing
  event Gaia16aye}. \aap 633:A98

\bibitem[{{Wyrzykowski} et~al.(2023){Wyrzykowski}, {Kruszy{\'n}ska}, {Rybicki},
  {Holl}, {Lec{\oe}ur-Ta{\"\i}bi}, {Mowlavi}, {Nienartowicz}, {Jevardat de
  Fombelle}, {Rimoldini}, {Audard}, {Garcia-Lario}, {Gavras}, {Evans},
  {Hodgkin}, and {Eyer}}]{wyrzykowski2023}
{Wyrzykowski} {\L}, {Kruszy{\'n}ska} K, {Rybicki} KA et~al. (2023) {Gaia Data
  Release 3. Microlensing events from all over the sky}. \aap 674:A23

\bibitem[{{Yan} and {Zhu}(2022)}]{yan2022}
{Yan} S {Zhu} W (2022) {Measuring Microlensing Parallax via Simultaneous
  Observations from Chinese Space Station Telescope and Roman Telescope}.
  Research in Astronomy and Astrophysics 22(2):025006

\bibitem[{{Yee} et~al.(2015){Yee}, {Gould}, {Beichman}, {Calchi Novati},
  {Carey}, {Gaudi}, {Henderson}, {Nataf}, {Penny}, {Shvartzvald}, and
  {Zhu}}]{yee2015b}
{Yee} JC, {Gould} A, {Beichman} C et~al. (2015) {Criteria for Sample Selection
  to Maximize Planet Sensitivity and Yield from Space-Based Microlens Parallax
  Surveys}. \apj 810(2):155

\bibitem[{{Yee} et~al.(2021){Yee}, {Zang}, {Udalski}, {Ryu}, {Green},
  {Hennerley}, {Marmont}, {Sumi}, {Mao}, {Gromadzki}, {Mr{\'o}z}, {Skowron},
  {Poleski}, {Szyma{\'n}ski}, {Soszy{\'n}ski}, {Pietrukowicz}, {Koz{\l}owski},
  {Ulaczyk}, {Rybicki}, {Iwanek}, {Wrona}, {Albrow}, {Chung}, {Gould}, {Han},
  {Hwang}, {Jung}, {Kim}, {Shin}, {Shvartzvald}, {Cha}, {Kim}, {Kim}, {Lee},
  {Lee}, {Lee}, {Park}, {Pogge}, {Bachelet}, {Christie}, {Hundertmark}, {Maoz},
  {McCormick}, {Natusch}, {Penny}, {Street}, {Tsapras}, {Beichman}, {Bryden},
  {Novati}, {Carey}, {Gaudi}, {Henderson}, {Johnson}, {Zhu}, {Bond}, {Abe},
  {Barry}, {Bennett}, {Bhattacharya}, {Donachie}, {Fujii}, {Fukui}, {Hirao},
  {Silva}, {Itow}, {Kirikawa}, {Kondo}, {Koshimoto}, {Alex Li}, {Matsubara},
  {Muraki}, {Miyazaki}, {Olmschenk}, {Ranc}, {Rattenbury}, {Satoh}, {Shoji},
  {Suzuki}, {Tanaka}, {Tristram}, {Yamawaki}, {Yonehara}, and {MOA
  Collaboration}}]{yee2021}
{Yee} JC, {Zang} W, {Udalski} A et~al. (2021) {OGLE-2019-BLG-0960 Lb: the
  Smallest Microlensing Planet}. \aj 162(5):180

\bibitem[{{Yoo} et~al.(2004){Yoo}, {DePoy}, {Gal-Yam}, {Gaudi}, {Gould}, {Han},
  {Lipkin}, {Maoz}, {Ofek}, {Park}, {Pogge}, {Mu-Fun Collaboration}, {Udalski},
  {Soszy{\'n}ski}, {Wyrzykowski}, {Kubiak}, {Szyma{\'n}ski}, {Pietrzy{\'n}ski},
  {Szewczyk}, {{\.Z}ebru{\'n}}, and {OGLE Collaboration}}]{yoo2004}
{Yoo} J, {DePoy} DL, {Gal-Yam} A et~al. (2004) {OGLE-2003-BLG-262:
  Finite-Source Effects from a Point-Mass Lens}. \apj 603(1):139--151

\bibitem[{{Zang} et~al.(2021){Zang}, {Hwang}, {Udalski}, {Wang}, {Zhu}, {Sumi},
  {Yee}, {Gould}, {Mao}, {Zhang}, {Albrow}, {Chung}, {Han}, {Jung}, {Ryu},
  {Shin}, {Shvartzvald}, {Cha}, {Kim}, {Kim}, {Kim}, {Lee}, {Lee}, {Lee},
  {Park}, {Pogge}, {Mr{\'o}z}, {Skowron}, {Poleski}, {Szyma{\'n}ski},
  {Soszy{\'n}ski}, {Pietrukowicz}, {Koz{\l}owski}, {Ulaczyk}, {Rybicki},
  {Iwanek}, {Wrona}, {Gromadzki}, {Bond}, {Abe}, {Barry}, {Bennett},
  {Bhattacharya}, {Donachie}, {Fujii}, {Fukui}, {Hirao}, {Itow}, {Kirikawa},
  {Kondo}, {Koshimoto}, {Li}, {Matsubara}, {Muraki}, {Miyazaki}, {Olmschenk},
  {Ranc}, {Rattenbury}, {Satoh}, {Shoji}, {Ishitani Silva}, {Suzuki}, {Tanaka},
  {Tristram}, {Yamawaki}, {Yonehara}, {Beichman}, {Bryden}, {Calchi Novati},
  {Carey}, {Gaudi}, {Henderson}, {Johnson}, and {Spitzer Team}}]{zang2021}
{Zang} W, {Hwang} KH, {Udalski} A et~al. (2021) {Systematic KMTNet Planetary
  Anomaly Search. I. OGLE-2019-BLG-1053Lb, a Buried Terrestrial Planet}. \aj
  162(4):163

\bibitem[{{Zang} et~al.(2022){Zang}, {Yang}, {Han}, {Lee}, {Udalski}, {Gould},
  {Mao}, {Zhang}, {Zhu}, {Albrow}, {Chung}, {Hwang}, {Jung}, {Ryu}, {Shin},
  {Shvartzvald}, {Yee}, {Cha}, {Kim}, {Kim}, {Kim}, {Lee}, {Lee}, {Park},
  {Pogge}, {Mr{\'o}z}, {Skowron}, {Poleski}, {Szyma{\'n}ski}, {Soszy{\'n}ski},
  {Pietrukowicz}, {Koz{\l}owski}, {Ulaczyk}, {Rybicki}, {Iwanek}, {Wrona}, and
  {Gromadzki}}]{zang2022a}
{Zang} W, {Yang} H, {Han} C et~al. (2022) {Systematic KMTNet planetary anomaly
  search. IV. Complete sample of 2019 prime-field}. \mnras 515(1):928--939

\bibitem[{{Zang} et~al.(2023){Zang}, {Jung}, {Yang}, {Zhang}, {Udalski}, {Yee},
  {Gould}, {Mao}, {Albrow}, {Chung}, {Han}, {Hwang}, {Ryu}, {Shin},
  {Shvartzvald}, {Cha}, {Kim}, {Kim}, {Kim}, {Lee}, {Lee}, {Lee}, {Park},
  {Pogge}, {KMTNet Collaboration}, {Mr{\'o}z}, {Skowron}, {Poleski},
  {Szyma{\'n}ski}, {Soszy{\'n}ski}, {Pietrukowicz}, {Koz{\l}owski}, {Ulaczyk},
  {Rybicki}, {Iwanek}, {Wrona}, {Gromadzki}, {OGLE Collaboration}, {Wang},
  {Zhang}, {Zhu}, and {MAP Collaboration}}]{zang2023a}
{Zang} W, {Jung} YK, {Yang} H et~al. (2023) {Systematic KMTNet Planetary
  Anomaly Search. VII. Complete Sample of $q < 10^{-4}$ Planets from the First
  4 yr Survey}. \aj 165(3):103

\bibitem[{{Zhu} and {Gould}(2016)}]{zhu2016}
{Zhu} W {Gould} A (2016) {Augmenting WFIRST Microlensing with a Ground-Based
  Telescope Network}. Journal of Korean Astronomical Society 49(3):93--107

\bibitem[{{Zhu} et~al.(2017){Zhu}, {Udalski}, {Novati}, {Chung}, {Jung}, {Ryu},
  {Shin}, {Gould}, {Lee}, {Albrow}, {Yee}, {Han}, {Hwang}, {Cha}, {Kim}, {Kim},
  {Kim}, {Kim}, {Lee}, {Park}, {Pogge}, {KMTNet Collaboration}, {Poleski},
  {Mr{\'o}z}, {Pietrukowicz}, {Skowron}, {Szyma{\'n}ski}, {KozLowski},
  {Ulaczyk}, {Pawlak}, {OGLE Collaboration}, {Beichman}, {Bryden}, {Carey},
  {Fausnaugh}, {Gaudi}, {Henderson}, {Shvartzvald}, {Wibking}, and {Spitzer
  Team}}]{zhu2017}
{Zhu} W, {Udalski} A, {Novati} SC et~al. (2017) {Toward a Galactic Distribution
  of Planets. I. Methodology and Planet Sensitivities of the 2015 High-cadence
  Spitzer Microlens Sample}. \aj 154(5):210

\end{thebibliography}

\end{document}